\documentclass[10pt,a4paper]{article}
\usepackage[dvips]{color}
\usepackage{epsfig}
\usepackage{epstopdf}
\usepackage{amsmath}
\usepackage{graphicx}
\usepackage{amssymb,amsmath}
\usepackage{amsmath}
\usepackage{tabularx}

\begin{document}
\title{\bf On a holographic dark energy model with a Nojiri-Odintsov cut-off in general relativity}
\author{{M. Khurshudyan$^{a}$\thanks{Email:khurshudyan@yandex.ru}}\\
$^{a}${\small {\em Institute for Physical Research, National Academy of Sciences, Ashtarak, Republic of Armenia}}\\
}\maketitle

\begin{abstract}
In this paper we consider the models of the accelerated expanding large scale universe~(according to general relativity) containing a generalized holographic dark energy with a Nojiri - Odintsov cut - off. The second component of the darkness is assumed to be the pressureless cold dark matter according to observed symmetries of the large scale universe. Moreover, we assume specific forms of the interaction between these two components and besides the cosmographic analysis, we discuss appropriate results from $Om$ and $Om3$ analysis and organize a closer look to the models via the statefinder hierarchy analysis, too. In this way we study mainly impact of the interaction on the dynamics of the background of our universe~(within specific forms of interaction). To complete the cosmographic analysis, the present day values of the statefinder parameters $(r,s)$ and $(\omega^{\prime}_{de}, \omega_{de})$ has been estimated for all cases and the validity of the generalized second law of thermodynamics is demonstrated. Our study showed that theoretical results from considered phenomenological models are consistent with the available observational data and symmetries.
\end{abstract}

\section{Introduction}\label{sec:INT}
The large scale universe has a long standing problem puzzling the researchers, which is known as the accelerated expansion of the large scale universe~\cite{Riess}~-~\cite{Perlmutter}. It is one of the central problems among the others. The minimal model to solve the problem is $\Lambda$CDM standard model of modern cosmology, where the cosmological constant plays the role of the dark energy. However, it is well known, that $\Lambda$CDM model poses two problems. Mainly, the value of the cosmological constant estimated from the observational data compared to the value estimated from quantum field theory, is very small~($10^{120}$ order). Even, if we change our believe to quantum field theory, we can reduce this difference up to $10^{50}$, which is still very big~\cite{Martin}~(and references therein). Surprisingly, mentioned problem can be avoided if we consider dynamical dark energy models~\cite{Yoo}~(and references therein). One of the first attempts to construct the dynamical dark energy models, was the replacement of the cosmological constant via the dynamical cosmological constant. This is very interesting research direction attracting appropriate attention in literature and, for instance, in Ref.~\cite{Maia} the reader can find some phenomenological forms of the varying cosmological constant model. Recently, various attempts to construct new models of the varying cosmological constant has been considered too~\cite{Sadeghi1}. Obtained results from the cosmographic analysis of suggested models do seem very promising. Another interesting class of the dark energy candidates, could be generated using the scalar fields. Among them are quintessence, phantom and quintom dark energy models~(to mention a few and see Ref.~\cite{Yoo} for an appropriate discussion). On the other hand, a special attention to the scalar field models of the dark energy is explained due to their applicability to the cosmic inflation puzzle. In history of our universe the cosmic inflation has a special place, because according to recent understanding, all required seeds to have our universe has been established namely during this accelerated expansion~\cite{Baumann}~(and references therein). There is an impressive amount of work, where the dark energy models with an explicitly given energy density have been involved to complete the darkness of the large scale universe. Among them is the generalized ghost dark energy~($\beta = 0$ corresponds to the ghost dark energy case)~\cite{Cai}
\begin{equation}\label{eq:GDE}
\rho_{de} = \alpha H + \beta H^{2},
\end{equation}  
where $\alpha$ and $\beta$ are constants, while $H$ it is the Hubble parameter of the universe. Recently, various phenomenological modifications of the ghost dark energy have been considered. Cosmographic analysis of the models revealed their viability and applicability to the problem of the accelerated expansion of the large scale universe according to the recent observational data~\cite{Khurshudyan}~-~\cite{Khurshudyan4}. Moreover, for considered cosmological models, with the varying ghost dark energy models, describing the large scale universe, have been demonstrate massless particle creation possibility in an appropriate radiation dominated early universes~\cite{Khurshudyan1}~-~\cite{Khurshudyan3}. This fact makes these models more attractive. More about technique behind massless particle creation in an expanding radiation dominated universe, reader can find in references of~\cite{Khurshudyan1}~-~\cite{Khurshudyan3}. Another option to represent the darkness of our universe it is the dark fluid. Actively discussed models of the dark energy fluids are barotropic dark fluid, polytropic dark fluid, Chaplygin gas and van der Waals gas~(with various modifications of mentioned fluids, including modifications via viscosity)~\cite{Khurshudyan5}~-\cite{Khurshudyan6}. Very often, we can meet various modifications and parameterizations of the EoS parameter of the dark fluid. A parameterization of the EoS parameter of the dark fluid involving the deceleration parameter $q$ has been considered recently in Ref~\cite{Khurshudyan7}. Seems that a parametrization of this kind is very natural, since origination of the dark energy in our universe will be directly related to the accelerated expansion according to such approach. Moreover, the opposite statement also could be correct. Another parameterization considered in Ref.~\cite{Khurshudyan7}, is related directly to the scalar model of the dark energy. The authors, have suggested to consider a scalar field model of the dark energy described by the following energy density and pressure
\begin{equation}
\rho_{\phi} = \frac{A}{2}\dot{\phi}^{2} + V(\phi), 
\end{equation}
\begin{equation}
P_{\phi} = \frac{A}{2}\dot{\phi}^{2} - V(\phi), 
\end{equation} 
which is kind of a joint model of the quintessence and phantom dark energy models, if we consider $A$ as a dynamical function defined on $[-1,1]$. Another interesting aspect in study of the accelerated expansion of the low redshift universe and the dark energy, it is various tools able to distinguish them among each other. One of such tools it is the statefinder analysis suggesting to study the following two parameters~\cite{Sahni} 
\begin{equation}
r = \frac{\dot{\ddot{a}}}{aH^{3}};
\end{equation}
and
\begin{equation}
s = \frac{r-1}{3(q-1/2)},
\end{equation}
where $a$ it is the scale factor of universe, while $q$ it is the deceleration parameter defined as
\begin{equation}
q = -1 - \frac{\dot{H}}{H^{2}}.
\end{equation}
It is s geometrical tool and has been applied successfully to many cosmological models since it has been proposed. Other possibilities allowing us to distinguish the dark energy models are $(\omega_{de}^{\prime}, \omega_{de})$~\cite{Caldwell}, $Om$~\cite{Sahni1}, $Om3$~\cite{Sahni3} and statefinder hierarchy analysis~\cite{Arabsalmani}. $(\omega_{de}^{\prime}, \omega_{de})$ analysis suggests to study behavior of the dark energy in $\omega_{de}^{\prime}$ and $\omega_{de}$ plane, where $\omega_{de}^{\prime}$ it is the derivative of the EoS of the dark energy with respect to $N=ln a$. $Om$ analysis suggests to study the following parameter
\begin{equation}
Om = \frac{x^{2}-1}{(1+z)^{3} - 1},
\end{equation}  
where $x = H/H_{0}$, $H$ it is the Hubble parameter and $H_{0}$ it is the value of the Hubble parameter at $z=0$. The three-point diagnostic $Om3$
\begin{equation}
Om3 = \frac{Om(z_{2},z_{1})}{Om(z_{3},z_{1})},
\end{equation}
where the two point $Om$ does read as
\begin{equation}
Om(z_{2},z_{1}) = \frac{x(z_{2})^{2} - x(z_{1}^{2})}{(1+z_{2})^{2} - (1+z_{1})^{2}},
\end{equation} 
is developed for mentioned purpose, too. On the other hand, the statefinder hierarchy analysis requires to calculate and study the following parameters 
\begin{equation}\label{eq:S3}
S^{(1)}_{3} = A_{3},
\end{equation}
\begin{equation}\label{eq:S4}
S^{(1)}_{4} = A_{4} + 3(1+q),
\end{equation}
\begin{equation}\label{eq:S5}
S^{(1)}_{5} = A_{5}  - 2 (4+ 3q)(1+q),
\end{equation}
etc., where $q$ it is the deceleration parameter, while $A_{n}$ reads as
\begin{equation}
A_{n} = \frac{a^{(n)}}{a H^{n}},
\end{equation} 
with 
\begin{equation}
a^{(n)} = \frac{d^{n}a}{dt^{n}}.
\end{equation}
Statefinder hierarchy for $\Lambda$CDM model during the cosmic expansion is equal to $1$. However, for the models with a dynamical dark energy and dark matter, $S^{(1)}_{n}$ are varying quantities and $\Lambda$CDM model can be chosen as a reference frame to emphasize possible deviations. Statefinder hierarchy analysis together with the growth rate study provides significantly important information about the model. Therefore, there is an active research to provide a closer look to the cosmological models via last two studies, too. Phase space analysis it is an elegant mathematical tool allowing to understand behavior of the low redshift universe. Moreover, it allows easily to find appropriate tracker solutions and demonstrate a solution to the cosmological coincidence problem. Phase space analysis removes needs to solve the field equations~\cite{Khurshudyan8}~(and references therein). Taking into account the importance of above mentioned tools, in this paper we will have a look to the cosmological models via $Om$ and $Om3$ analysis to complete the estimation of the present day values of $(r,s)$ and $(\omega_{de},\omega^{\prime}_{de})$ parameters, when there are specific forms of the interaction between the dark energy and dark matter. We will address the next section to the details of the cosmological models defined according to general relativity and in the same section we will include an appropriate discussion on the forms of the interaction term. The lack of the fundamental theory, still leaves the questions open concerning to the form and the reasons of having an interaction inside the darkness of the large scale universe. On the other hand, available data is not enough to have appropriate statistical analysis allowing to gain more information concerning to these questions. In literature, there is an appropriate discussion on this topic and we refer the reader to the references of this work for an appropriate information~\cite{Maia}~-~\cite{Khurshudyan8} and references therein. To finalize this section, we would like very briefly mention about modified theories of gravity. Modified theories of gravity are attractive, since they are trying to explain the observational data providing different background dynamics to universe. If we apply modified theories of gravity, we do not need to introduce the dark energy by hand as in case of general relativity, since an appropriate modification provides an additional term in the field equations, which can be interpreted as the dark energy. In this case, the origin of the dark energy it is in the heart of the modified theory, but each modifications provides its version of the dark energy. Therefore the modification of general relativity it is one step forwards towards the solutions of the large scale universe, but not the final one. This is the reason for looking for other radical ways to solve the problems~\cite{Nojiri}~-~\cite{Clifton}.\\\\
The paper is organized as follows: In section~\ref{sec:INTMGR} we will present a detailed description of the suggested cosmological models containing interacting generalized holographic dark energy model with a Nojiri - Odintsov cut - off. In section~\ref{sec:INTM} we will present and discuss results from the cosmographic analysis of the models involving the estimation of the present day values of the statefinder parameters $(r,s)$ and $(\omega^{\prime}_{de}, \omega_{de})$ for different forms of interaction. In section~\ref{sec:Om} we will study the models involving $Om$ and $Om3$ analysis at the same time organizing a look to the models via the statefinder hierarchy analysis. Moreover, in section~\ref{sec:ST} the validity of the generalized second law of thermodynamics is demonstrated. Finally, a discussion on the obtained results and possible future extension of considered cosmological models are summarized in section~\ref{sec:Discussion}.

\section{Interacting dark energy models in General Relativity}\label{sec:INTMGR}
There is a simple framework to describe interacting dark energy models, when we consider general relativity to describe the dynamics of the background. In this paper, we consider cosmological models, which could be applicable to the accelerated expansion of the large scale universe, therefore two fluids approximation will be used. According to this approximation, the darkness of the large scale universe can be described by an effective fluid with the energy density and pressure defined as
\begin{equation}\label{eq:rhoeff}
\rho_{eff} = \rho_{de} + \rho_{dm},
\end{equation} 
\begin{equation}\label{eq:Peff}
P_{eff} = P_{de} + P_{dm},
\end{equation}
where $\rho_{de}$ and $\rho_{dm}$ are the energy densities, while $P_{de}$ and $P_{dm}$ are the pressures of the dark energy and dark matter, respectively. With this assumption, the dynamics of the energy densities will take the following form
\begin{equation}\label{eq:rhoDE}
\dot{\rho}_{de} + 3 H \rho_{de}(1+\omega_{de}) = -Q,
\end{equation} 
\begin{equation}\label{eq:rhoDM}
\dot{\rho}_{dm} + 3 H \rho_{dm} = Q,
\end{equation} 
where $Q$ stands for the interaction inside the darkness, $\omega_{de}$ it is the EoS parameter of the dark energy. while the energy density of the effective fluid, Eq.~(\ref{eq:rhoeff}), allows to determine the Hubble parameter $H$~(in $8 \pi G = c = 1$ units)
\begin{equation}\label{eq:Hubble}
H^{2} = \frac{1}{3} \rho_{eff}.
\end{equation}
The structure of the equations describing the dynamics of the large scale universe, Eq.~(\ref{eq:rhoDE})~-~(\ref{eq:Hubble}), demand additional assumptions concerning either about the form of the interaction $Q$ and the Hubble parameter $H$~(it could be also the scale factor $a$), or the form of the interaction term $Q$ and the EoS of the dark energy. Another option is to consider specific form for the $H$ and the EoS, and obtain the form/constraints on the interaction term $Q$. In this paper, we will suggest the forms of the interaction term $Q$ and the EoS of the dark energy, particularly, we will consider a cosmological model, where the interaction inside the darkness is given by 
\begin{equation}\label{eq:Q1}
Q =  3 H b (\rho_{de} + \rho_{dm}),
\end{equation}
and compare this model with the models, where a certain class of nonlinear interactions are involved. Considered nonlinear interactions will be obtained from the following general form of the interaction for $m=0$
\begin{equation}\label{eq:QGF}
Q = 3 H b q^{m} \frac{\rho_{i}\rho_{j}}{\rho_{de} + \rho_{dm}},
\end{equation}     
where $b$ and $m$ are positive constants, while $\rho_{i}$ and $\rho_{j}$ stands for the energy densities either of the dark energy or the dark matter. The form of the interaction given by Eq.~(\ref{eq:QGF}), contains two types of nonlinear interactions. Particularly, when $m=1$, we have the sign changeable nonlinear interaction, which has been achieved due to the deceleration parameter $q$.
There is an active discussion on the sign changeable interaction between the dark energy and dark matter. Particularly, there is an increasing interest towards to the sign changeable interactions of different nature~(see for instance Ref.~\cite{Khurshudyan9}). On the other, concerning to the EoS of the dark energy we will follow to Ref~\cite{Nojiri2}. and we will consider a particular model of the generalized holographic dark energy with the Nojiri- Odintsov cut - off defined as
\begin{equation}\label{eq:NOHDE}
\rho_{de} = \frac{3c^{2}}{L^{2}},
\end{equation}
with
\begin{equation}\label{eq:CUTTOFF}
\frac{c}{L} = \frac{1}{L_{f}} \left [ \alpha_{0} + \alpha_{1} L_{f} + \alpha_{2} L_{f}^{2}\right ] 
\end{equation}
where $L_{f}$ it is the future horizon and defined as 
\begin{equation}\label{eq:LF}
L_{f} = a\int_{t}^{\infty}{\frac{dt}{a}},  
\end{equation}
while $c$, $\alpha_{0}$, $\alpha_{1}$ and $\alpha_{2}$ are numerical constants. About a detailed discussion concerning to the reasons of consideration of such cut-offs can be found in Ref.~\cite{Nojiri2}, where it is demonstrated a possibility of the unifying of the early -time and late - time universe based on phantom cosmology. Moreover, one of the interesting results~(among the others) discussed in Ref.~\cite{Nojiri2} is related to the possibility of phantom -- non-phantom transition, which appears in such a way that the universe could have effectively phantom equation of state at early - time as well as at late - time. Generally, the oscillating universe may have several phantom and non-phantom phases. On the other hand, we would like to mention about Ref.~\cite{Nojiri2}, where another example of the cut - off has been suggested and considered. The form of the cut - off given by Eq.~(\ref{eq:CUTTOFF}) can be understood as a particular example of the general form
\begin{equation}\label{eq:CUTTOFF}
\frac{c}{L} = \frac{1}{L_{f}} \sum_{i}{\alpha_{i} L_{f}^{i}}.
\end{equation}
It is not hard to see that $\dot{L}_{f} = H L_{f} - 1$, which will be used in future during the presentation of the results in appropriate sections. We start our discussion from the next section starting from the cosmography of the model.

\section{Cosmography of the models}\label{sec:INTM}
To simplify our discussion on the results from the cosmographic analysis we organized two subsections. We start our analysis from the model, where the interaction between the dark energy and dark matter is given by Eq.~(\ref{eq:Q1}). For sake of simplicity we have imposed $\alpha_{0} \in [0,1]$, $\alpha_{1}\in [0,1]$ and $\alpha_{2} \in [0,1]$ constraints on the parameters of the dark energy model. Moreover, the best fit of the theoretical results with the distance modulus has been used to organize the discussion having more precise constraints on $\alpha_{0}$, $\alpha_{1}$, $\alpha_{2}$ and on the interaction parameter $b$. This allows us to save appropriate space. 

\subsection{Models with $Q = 3Hb (\rho_{de} + \rho_{dm})$}
\begin{figure}[h!]
 \begin{center}$
 \begin{array}{cccc}
\includegraphics[width=80 mm]{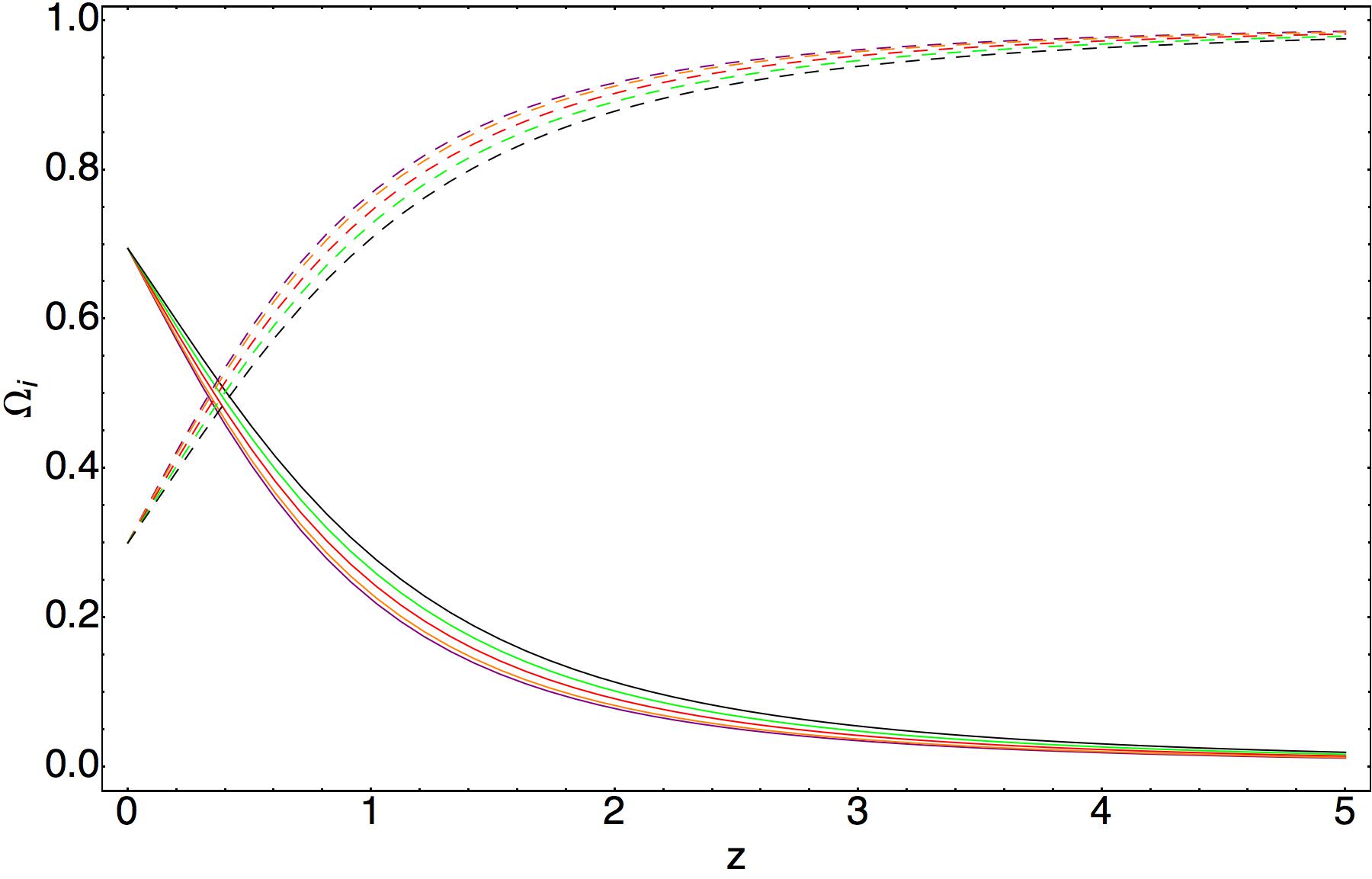} &
\includegraphics[width=78 mm]{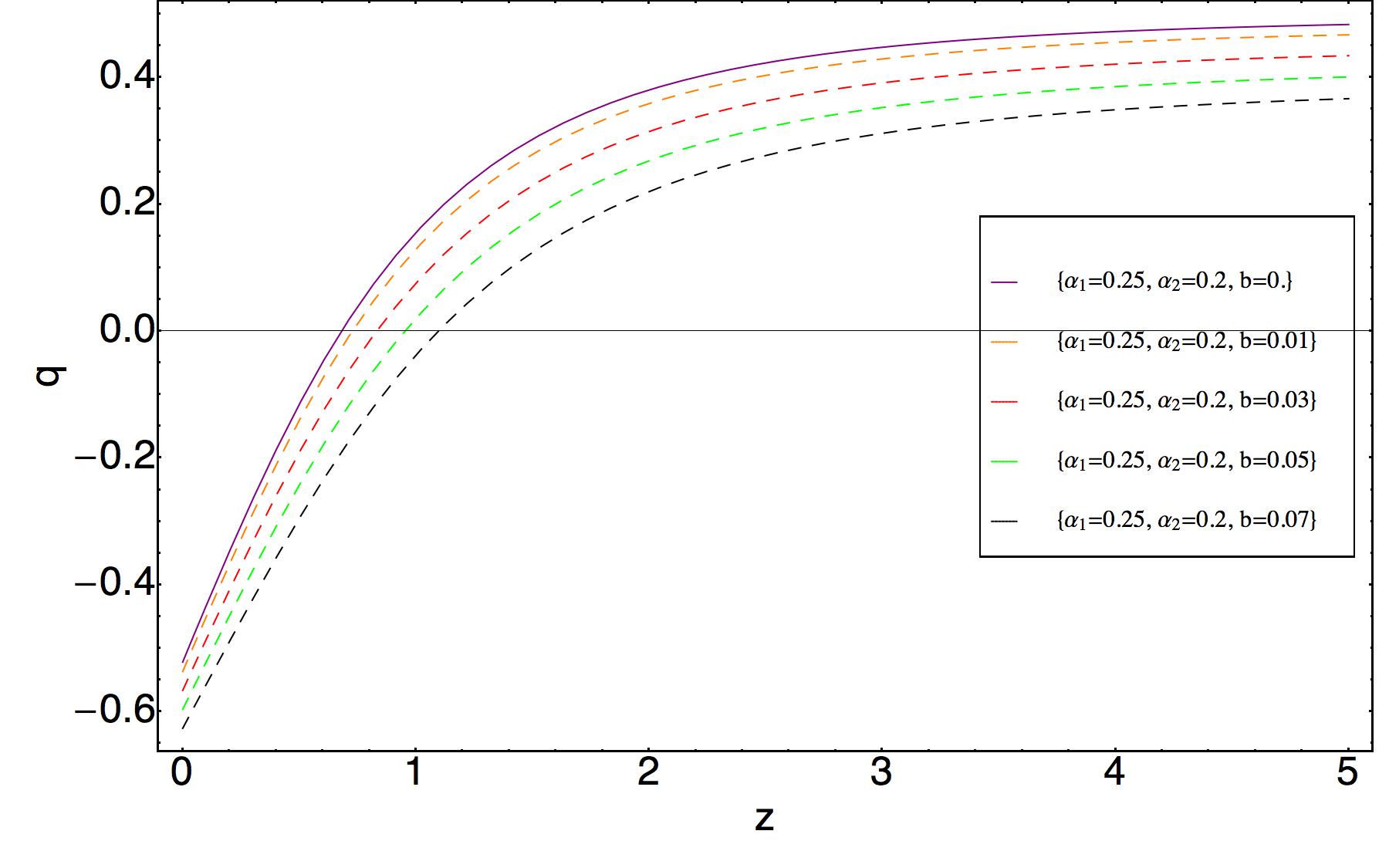}  
 \end{array}$
 \end{center}
\caption{Graphical behavior of the deceleration parameter $q$, $\Omega_{de}$ and $\Omega_{dm}$ against the redshift $z$ for the cosmological model, where the interaction between the dark energy and dark matter is given by Eq.~(\ref{eq:q1}). Presented behavior of for $\Omega_{de}$ and $\Omega_{dm}$ is according to the same values of the parameters as for the behavior of the deceleration parameter $q$. The solid lines on $\Omega_{i} - z$ plane represent the behavior of $\Omega_{de}$, while dashed lines represent the behavior of $\Omega_{dm}$. Considered model is free from the cosmological coincidence problem.}
 \label{fig:Fig1}
\end{figure}

\begin{figure}[h!]
 \begin{center}$
 \begin{array}{cccc}
\includegraphics[width=80 mm]{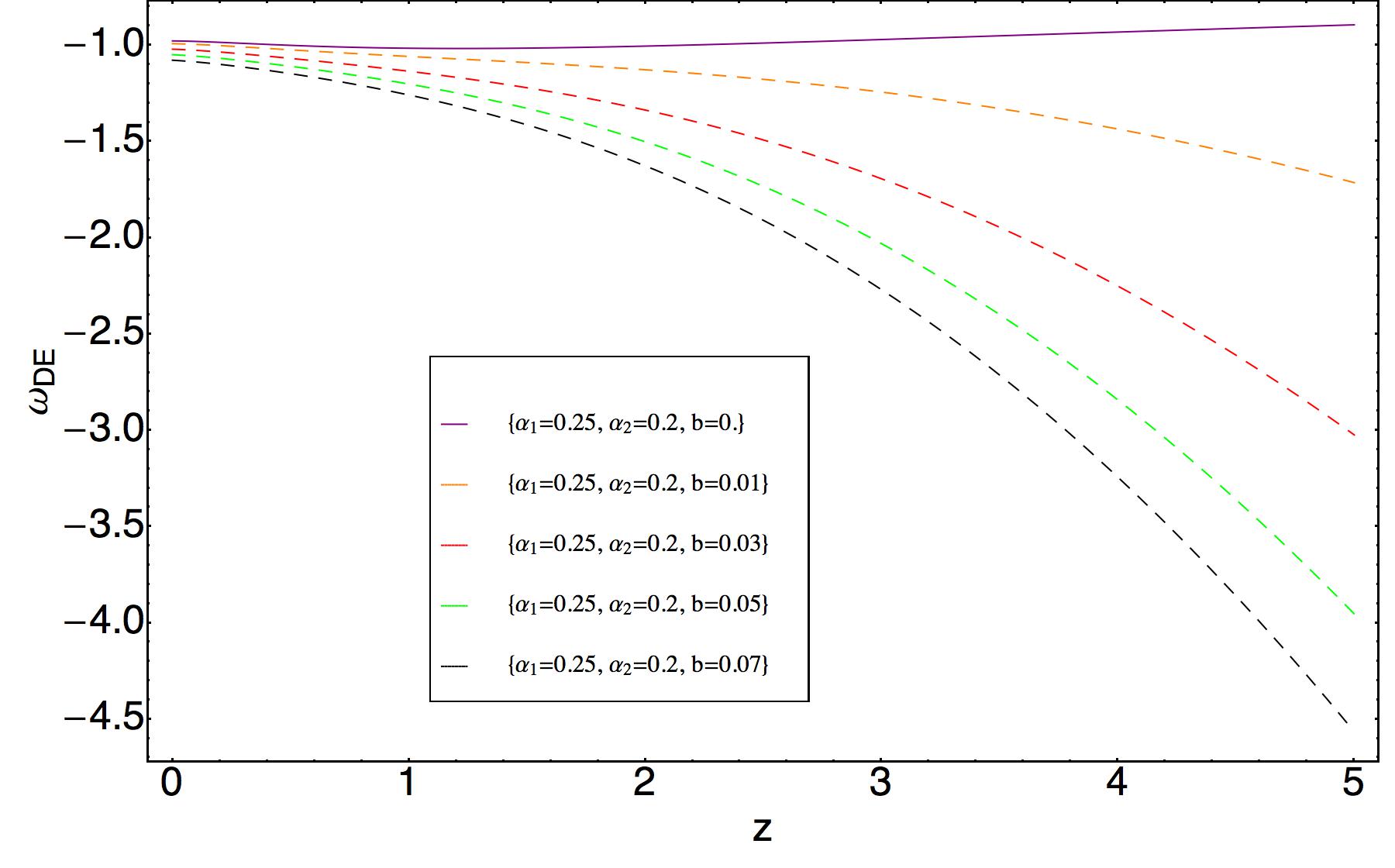} &
\includegraphics[width=80 mm]{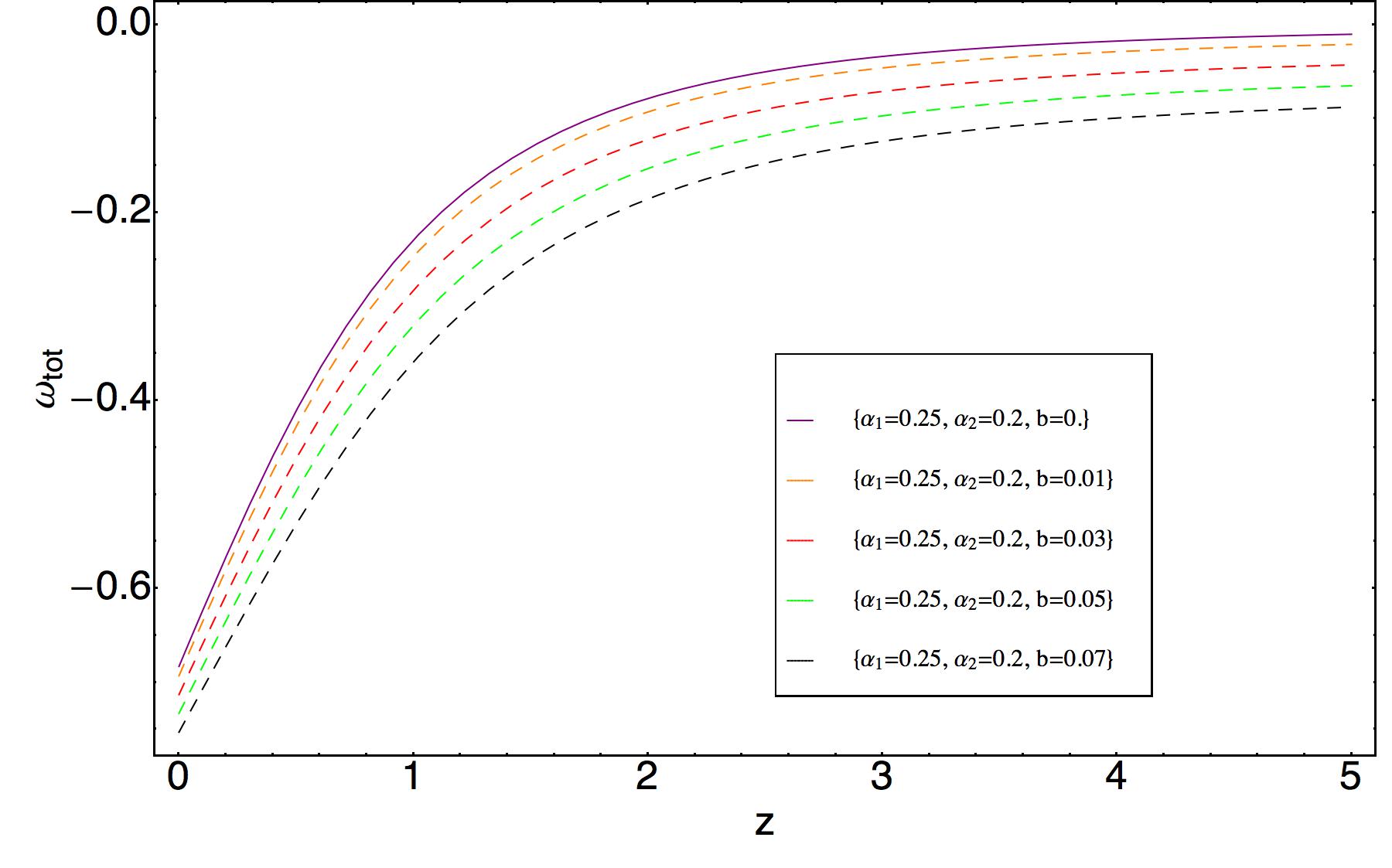} \\
 \end{array}$
 \end{center}
\caption{Graphical behavior of the EoS parameter of the dark energy against the redshift $z$ represents the left plot. The right plot represents the behavior of the EoS parameter of the effective fluid. Considered case corresponds to the model, where the interaction between the dark energy and dark matter is given by Eq.~(\ref{eq:q1}). For higher redshifts the effective fluid is a fluid with $\omega_{tot} > 0$.}
 \label{fig:Fig2}
\end{figure}

The cosmological model of the large scale universe, where the interaction between the dark energy and dark matter is given by Eq.~(\ref{eq:Q1}), contains the dark energy with the following EoS parameter
\begin{equation}
\omega_{de} = -\frac{3 b H^2 L_{f} + 2 \sqrt{\Omega _{de}} \dot{L}_{f} (\alpha_{1} + 2 \alpha_{2} L_{f}) + H \Omega _{de} (H L_{f} + 2 ) } {3 H^2 L_{f} \Omega _{de}},
\end{equation}
while the deceleration parameter $q$ reads as
\begin{equation}\label{eq:q1}
q = \frac{(1-3 b) H^2 L_{f}-2 \sqrt{\Omega _{de}} \dot{L}_{f} (\alpha_{1} + 2 \alpha_{2} L_{f})-H \Omega _{de} (H L_{f}+2)}{2 H^2 L_{f}}.
\end{equation}
This is a model of the large scale universe, where the phase transition to the accelerated expanding universe took place for $z_{tr} \approx 0.682$ when $b = 0$. Recall, that non interacting model according to presented setup of the models corresponds to $Q=0$. On the other hand, $b=0$ condition leads to $Q=0$, therefore, the non interacting model in our case corresponds to $b=0$ case. On the other hand, increasing the value of the interaction parameter $b$ will bring to an increase of the transition redshift $z_{tr}$ with an appropriate decrease of the present day value of the deceleration parameter $q$ defined by Eq.~(\ref{eq:q1}). The impact of the interaction, Eq.~(\ref{eq:Q1}), can be found imprinted into the dynamics of $\Omega_{de}$ and $\Omega_{dm}$. Compared to the non interacting model, an increase of the interaction parameter $b$ will speed up an increase of the amount of $\Omega_{de}$ in our universe giving an appropriate decrease of $\Omega_{dm}$. However, this model is free from the cosmological coincidence problems and presented information about the behavior of the deceleration parameter, $\Omega_{de}$ and $\Omega_{dm}$ can be found in Fig.~(\ref{fig:Fig1}). Moreover, we have studied the behavior of the EoS parameter of the dark energy and the EoS parameter of the effective fluid. As it can be seen from the right plot of Fig.~(\ref{fig:Fig2}), the EoS of the effective fluid indicates a quintessence large scale universe. On the other hand, we see from the same plot, that an increase of the value of $b$ will decrease the EoS parameter of the effective fluid. An interesting behavior has been observed for the EoS parameter of the dark energy, which is presented on the left plot of the Fig.~(\ref{fig:Fig2}). Particularly, we see that for the recent epoch the value of the EoS parameter of the dark energy is within the range of recently obtained constraints from the Planck satellite 2015 experiments for $b \in [0,0.03]$~\cite{Planck}. Moreover, an increase of $b$ gives an increase of $\omega^{\prime}_{de}$ and a decrease of $\omega_{de}$ for the lower redshifts. On the other hand, for the higher redshifts we observed, that with an increase of the value of the parameter $b$ the nature of the EoS has been changed from the quintessence to the phantom. If we demand only the quintessence nature for the dark energy during the evolution of the universe, then we can obtain another constraint on the parameter $b$, namely, we will have $b \in [0, 0.005]$. The present day values of the deceleration parameter $q$, $(\omega_{de},\omega^{\prime}_{de})$ of interacting dark energy, $(r,s)$ statefinder parameters and the value of the transition redshift for some values of the interaction parameter $b$ are presented in Table~\ref{tab:Table1}. We see from Table~\ref{tab:Table1} that increase of the parameter $b$ as it is discussed for the graphical behavior of cosmological parameters, will decrease the present day value of $r$ and will increase the present day value of $s$ parameters.

\begin{table}[ht!]
  \centering
    \begin{tabular}{ | l | l | l | l | p{1cm} |}
    \hline
 $b$ & $q$ & $(\omega_{de},\omega^{\prime}_{de})$ & $(r,s)$ & $z_{tr}$ \\
      \hline
 $0.0 $ & $-0.522$ & $(-0.978, 0.249)$ & $(2.86,  -0.61)$ & $0.682$ \\
          \hline
 $0.01$ & $-0.537$ & $(-0.993, 0.257)$ & $(2.76,  -0.57)$ & $0.728$ \\
    \hline
 $0.03$ & $-0.567$ & $(-1.022, 0.269)$ & $(2.57,  -0.49)$ & $0.832$ \\
     \hline
 $0.05$ & $-0.596$ & $(-1.051, 0.279)$ & $(2.39,  -0.42)$ & $0.954$ \\
     \hline
 $0.07$ & $-0.626$ & $(-1.079, 0.284)$ & $(2.23,  -0.36)$ & $1.101$ \\
     \hline            
    \end{tabular}
\caption{Present day values of the deceleration parameter $q$, $(\omega_{de},\omega^{\prime}_{de})$ of interacting dark energy, $(r,s)$ statefinder parameters and the value of the transition redshift $z_{tr}$ for several values of the interaction parameter $b$, when the interaction is given via Eq.~(\ref{eq:Q1}). The best fit of the theoretical results to the recent observational data has been obtained for $H_{0}=0.7$, $\alpha_{0} = 0.15$, $\alpha_{1} = 0.25$, $c = 0.75$.}
  \label{tab:Table1}
\end{table}

\subsection{Models with non linear interactions}
\begin{figure}[h!]
 \begin{center}$
 \begin{array}{cccc}
\includegraphics[width=80 mm]{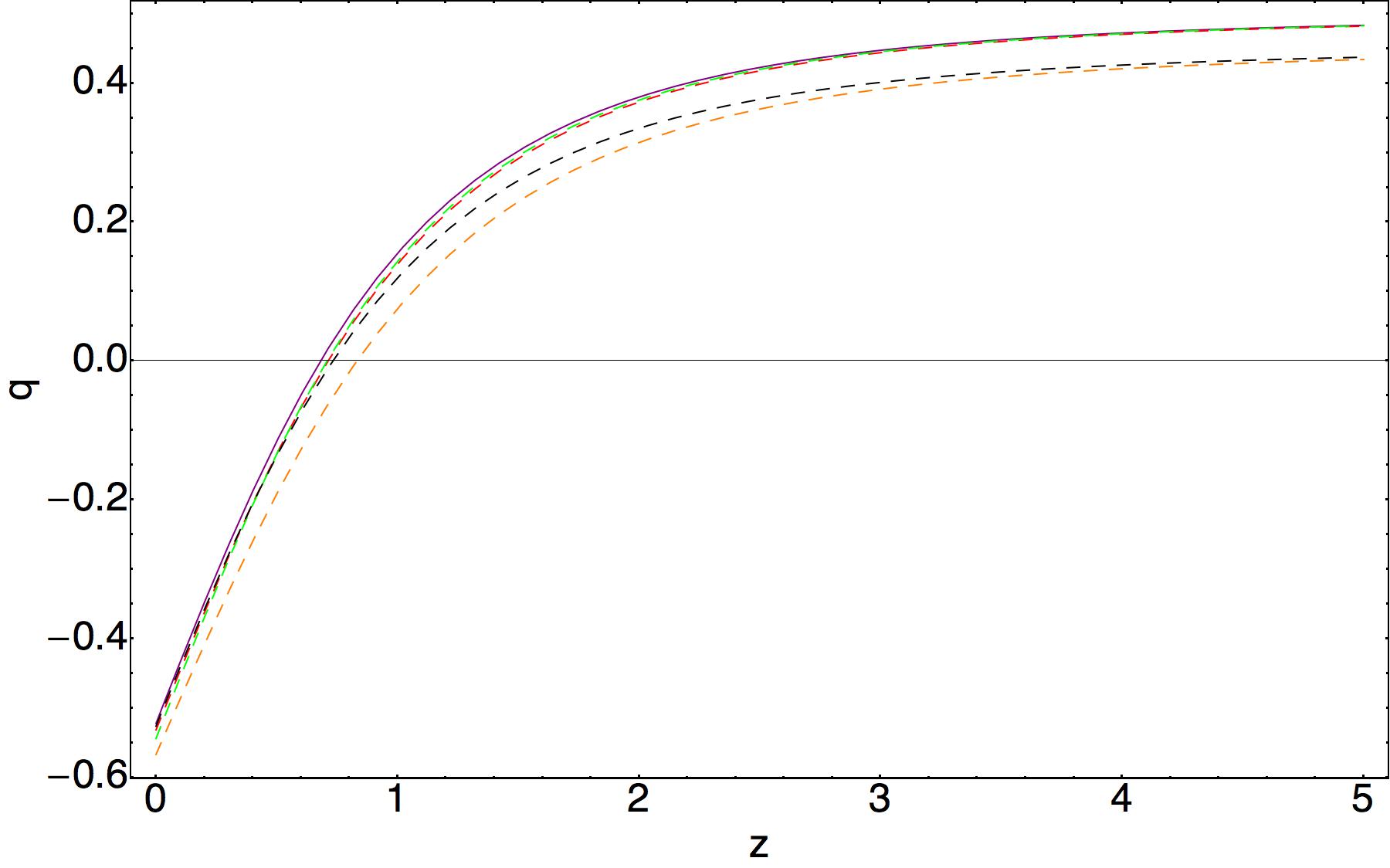} &
\includegraphics[width=78 mm]{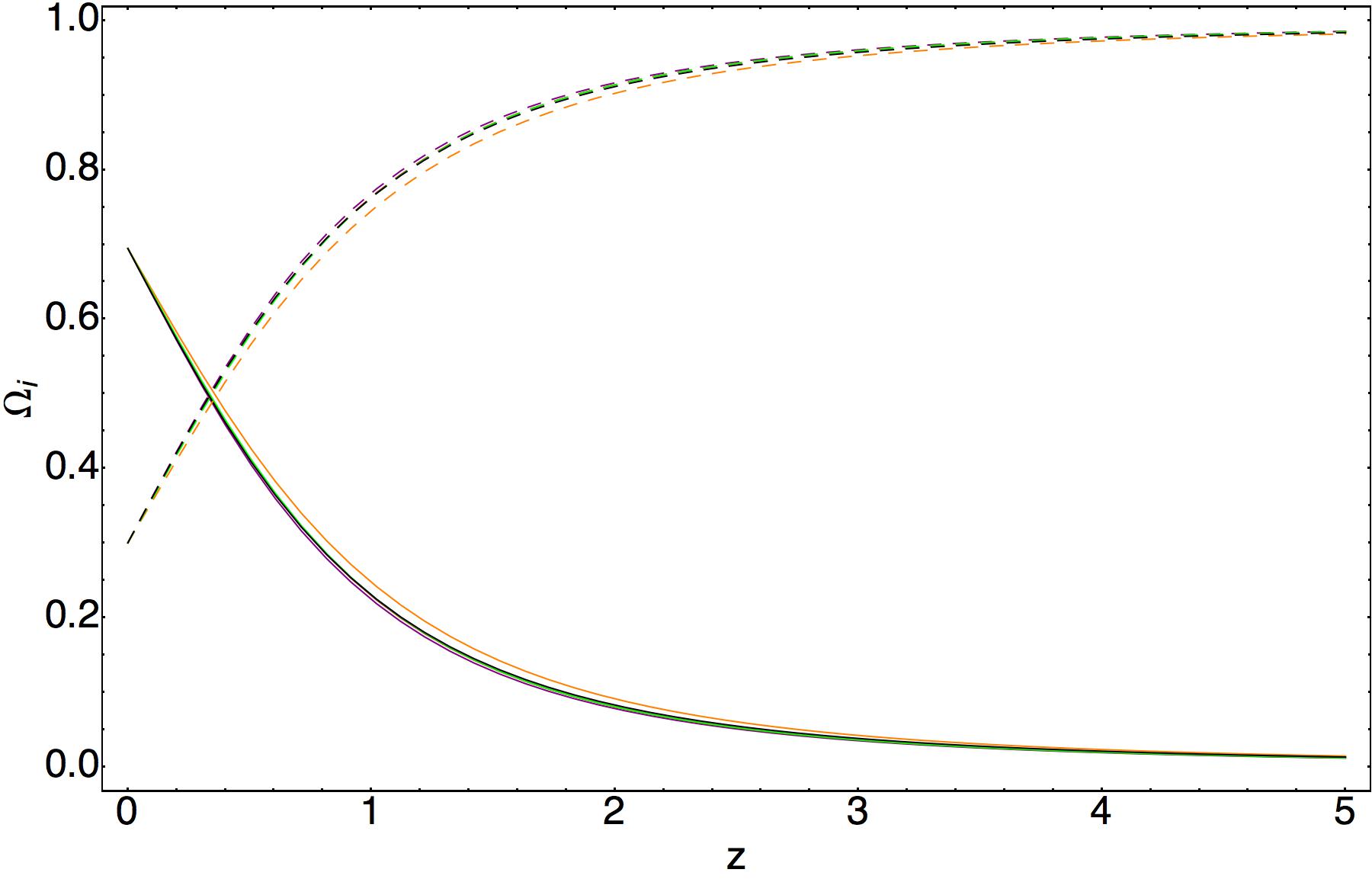} \\
\includegraphics[width=80 mm]{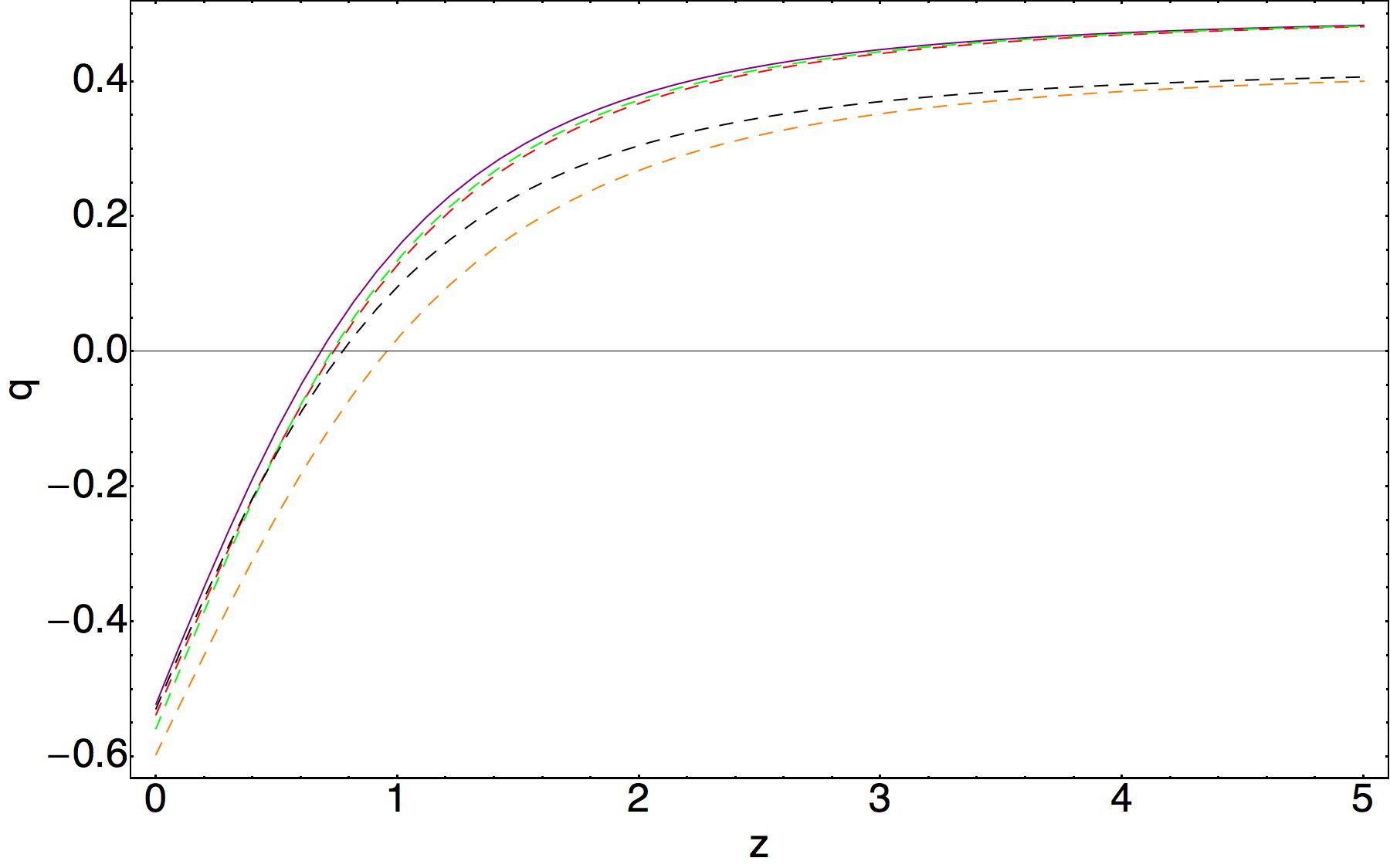} &
\includegraphics[width=78 mm]{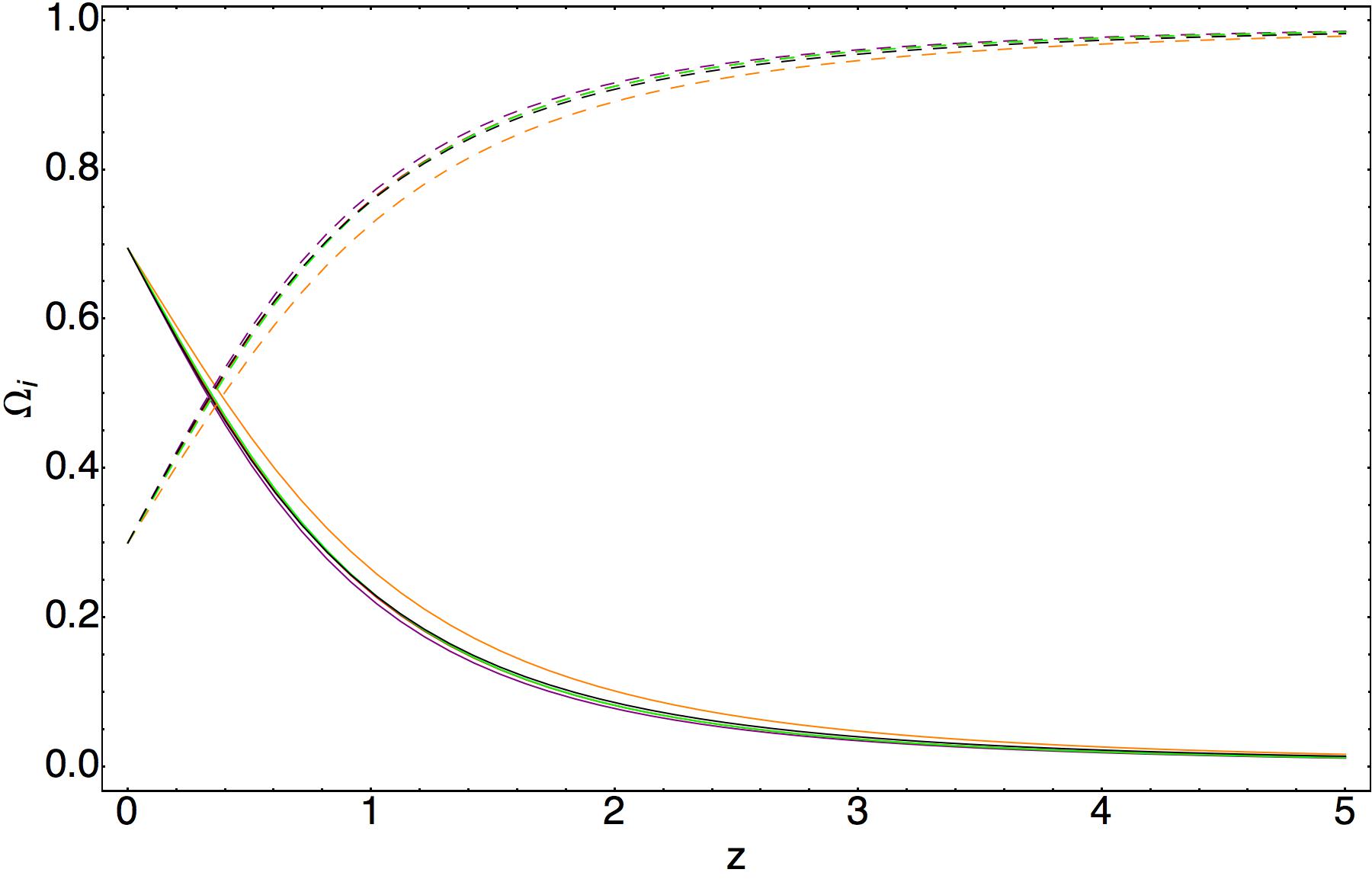} \\
 \end{array}$
 \end{center}
\caption{Graphical behavior of the deceleration parameter $q$, $\Omega_{de}$ and $\Omega_{dm}$ against the redshift $z$. The blue curve represents non interacting model with $b = 0$, the orange curve represents the model, when the interaction is given by Eq.~(\ref{eq:Q1}), the red curve represents the model when the interaction is given by Eq.~(\ref{eq:Q2}), the green curve represents the model with the interaction given by Eq.~(\ref{eq:Q3}), while the black curve represents the model with the interaction given by Eq.~(\ref{eq:Q4}), when $H_{0}=0.7$, $\alpha_{0} = 0.15$, $\alpha_{1} = 0.25$, $c = 0.75$. The top panel corresponds to the case when $b = 0.03$. The bottom panel represents the case when for the interacting models $b=0.05$.}
 \label{fig:Fig3}
\end{figure}

\begin{figure}[h!]
 \begin{center}$
 \begin{array}{cccc}
\includegraphics[width=80 mm]{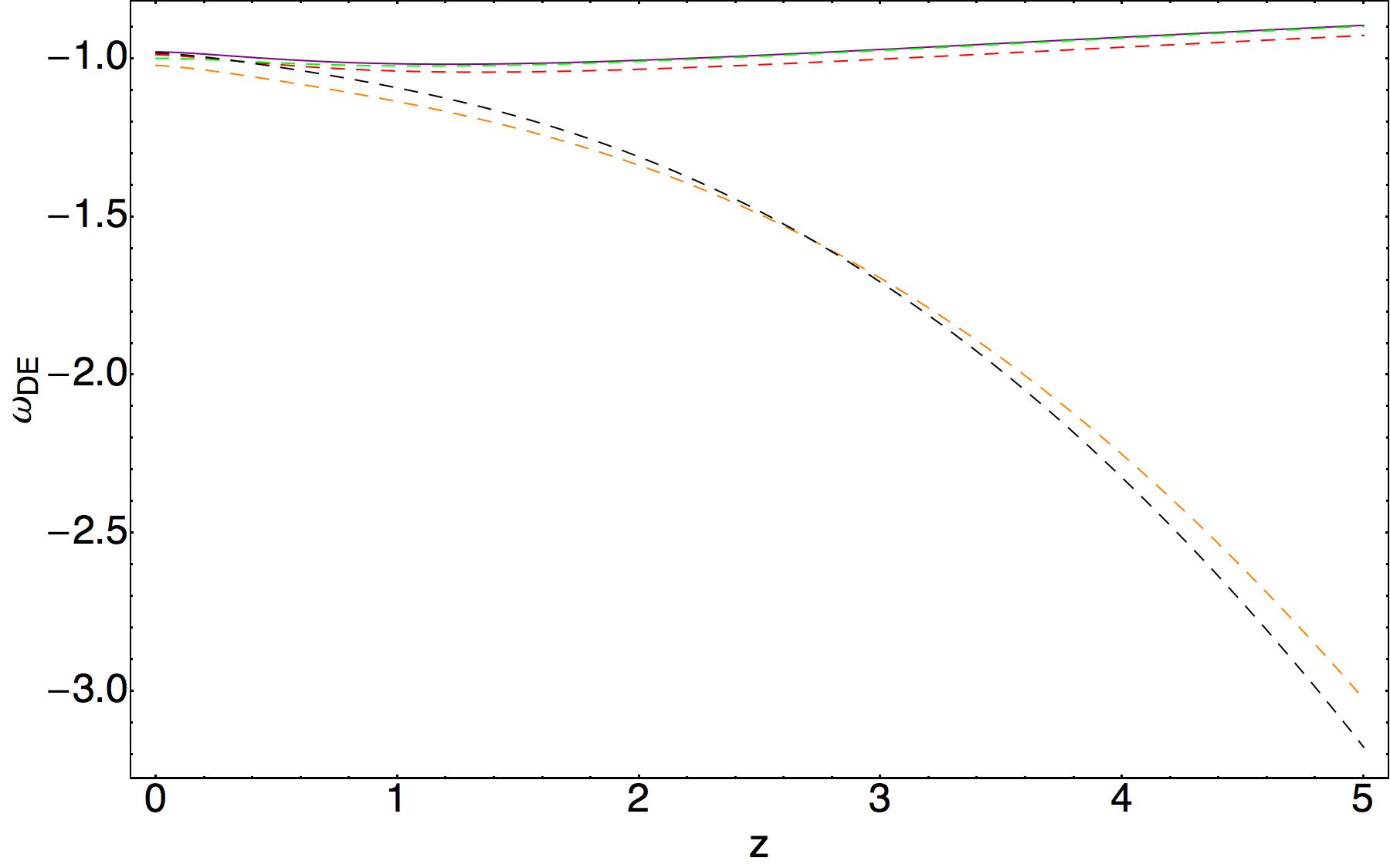} &
\includegraphics[width=78 mm]{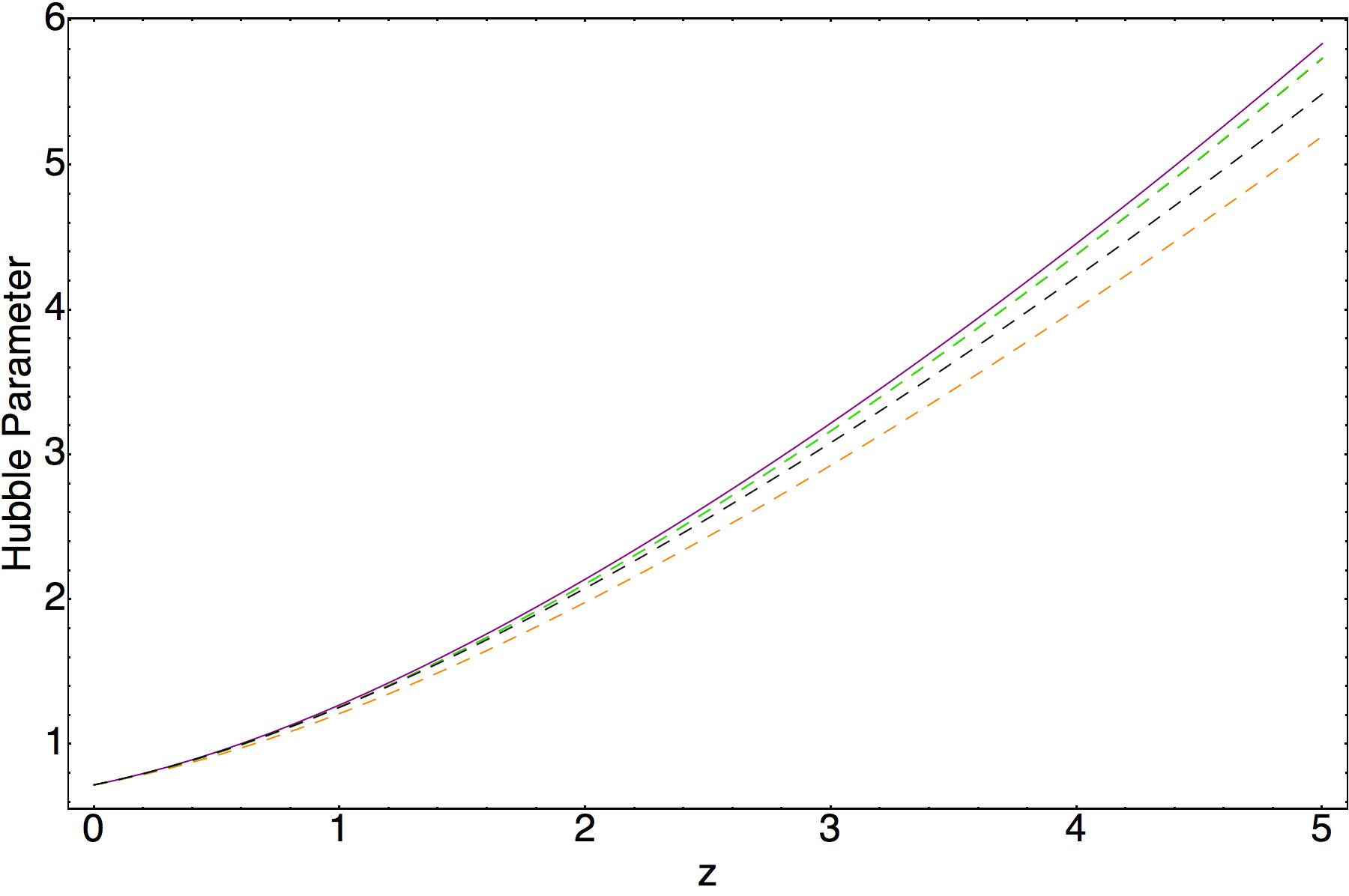} \\
\includegraphics[width=80 mm]{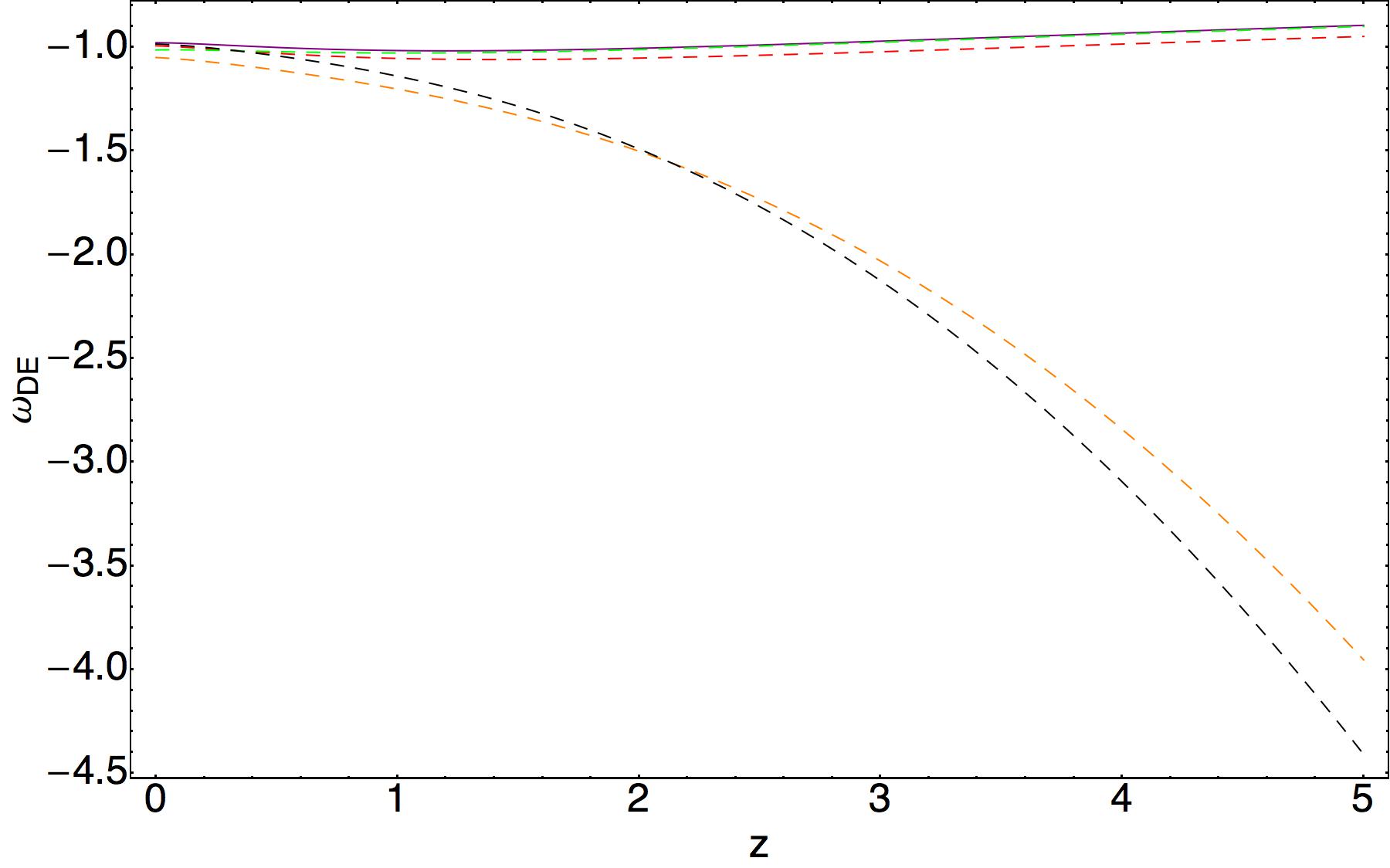} &
\includegraphics[width=78 mm]{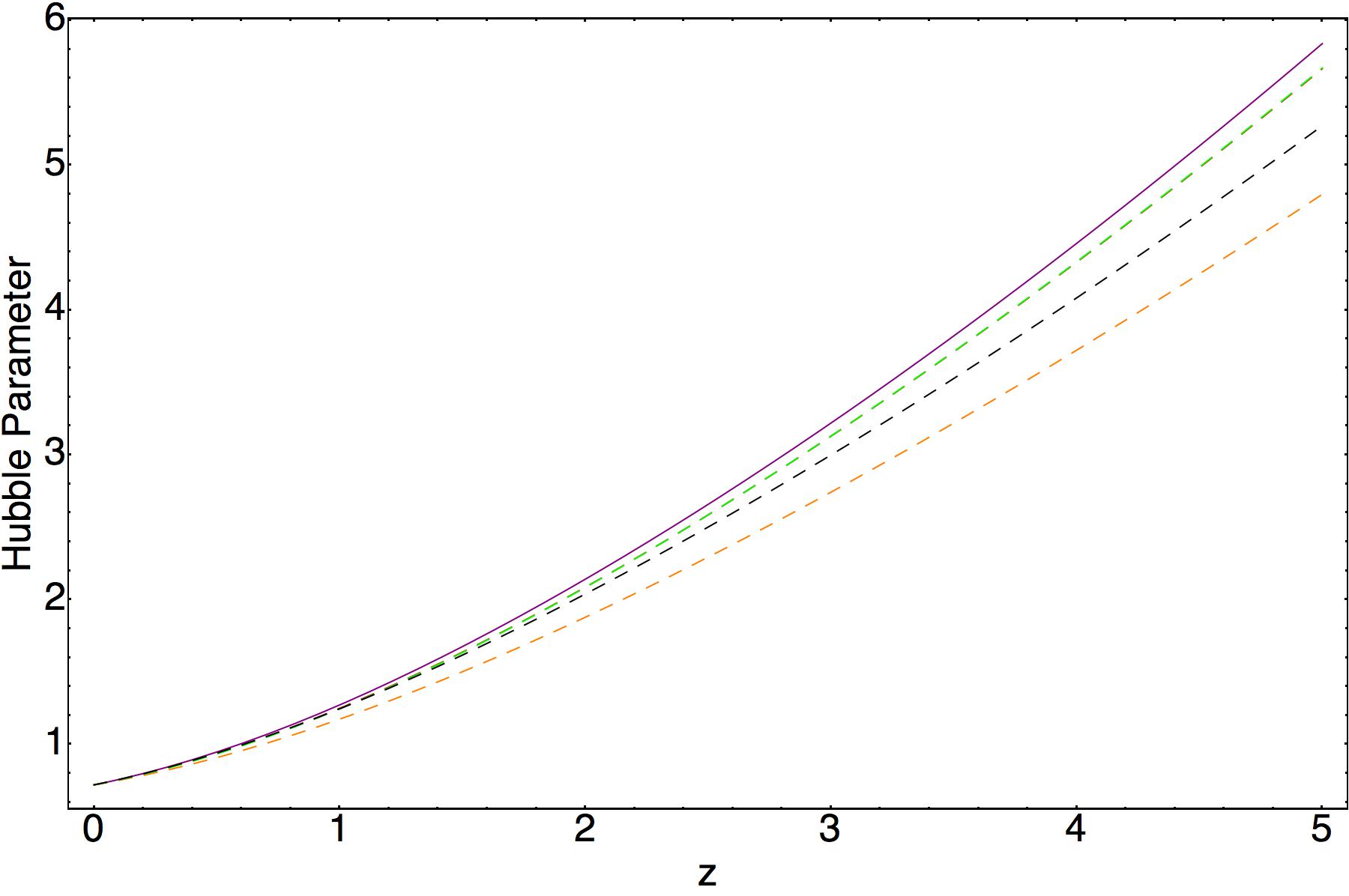} \\
 \end{array}$
 \end{center}
\caption{Graphical behavior of the EoS parameter $\omega_{de}$ of the dark energy and the Hubble parameter $H$ against the redshift $z$. The blue curve represents non interacting model with $b = 0$, the orange curve represents the model, when the interaction is given by Eq.~(\ref{eq:Q1}), the red curve represents the model when the interaction is given by Eq.~(\ref{eq:Q2}), the green curve represents the model with the interaction given by Eq.~(\ref{eq:Q3}), while the black curve represents the model with the interaction given by Eq.~(\ref{eq:Q4}), when $H_{0}=0.7$, $\alpha_{0} = 0.15$, $\alpha_{1} = 0.25$, $c = 0.75$. The top panel corresponds to the case when $b = 0.03$. The bottom panel represents the case when for the interacting models $b=0.05$.}
\label{fig:Fig4}
\end{figure}

In this subsection we will consider the models, where the interaction involved into the darkness of the large scale universe has been obtained from Eq.~(\ref{eq:QGF}), when $m=0$. One of the examples of such interaction is 
\begin{equation}\label{eq:Q2}
Q = 3 b H  \frac{\rho_{de} \rho_{dm}}{\rho_{de} + \rho_{dm}},
\end{equation}
which gives an universe, where the EoS parameter of the dark energy and the deceleration parameter in terms of $\Omega_{de}$, $\alpha_{1}$, $\alpha_{2}$ and $b$ read as
\begin{equation}
\omega_{de} = \frac{1}{3} \left(3 b \Omega_{de}-3 b-\frac{2 \dot{L}_{f} (\alpha_{1} + 2 \alpha_{2} L_{f})}{H^2 L_{f} \sqrt{\Omega_{de}}}-\frac{2}{H L_{f}}-1\right)
\end{equation}
and
\begin{equation}
q = \frac{3 b H^2 L_{f} \Omega_{de}^2-H \Omega _{de} (H (3 b L_{f}+ L_{f})+2)-2 \sqrt{\Omega _{de}} \dot{L}_{f} (\alpha_{1} + 2 \alpha_{2} L_{f}) + H^2 L_{f}}{2 H^2 L_{f}}.
\end{equation}
Fig.~(\ref{fig:Fig3}) represents the graphical behavior of the deceleration parameter $q$, $\Omega_{de}$ and $\Omega_{dm}$ for the cosmological models where the interaction is given by Eq.~(\ref{eq:Q1}), Eq.~(\ref{eq:Q2}),
\begin{equation}\label{eq:Q3}
Q = 3 b H  \frac{\rho^{2}_{de}}{\rho_{de} + \rho_{dm}},
\end{equation}
and
\begin{equation}\label{eq:Q4}
Q = 3 b H  \frac{\rho^{2}_{dm}}{\rho_{de} + \rho_{dm}},
\end{equation}
and simplify our discussion. From  Fig.~(\ref{fig:Fig3}) we see, that for the same values of the parameters of the models, consideration of the interaction, Eq.~(\ref{eq:Q1}), compared to non interacting case with $b=0$ and other forms of non linear interactions, provides the highest value for the transition redshift $z_{tr}$~(orange curve). Moreover, other forms of interactions, Eq.~(\ref{eq:Q2}), Eq.~(\ref{eq:Q3}) and Eq.~(\ref{eq:Q4}), will also increase the transition redshift. On the other hand, an appropriate increase of the transition redshift will decrease the present day value of the deceleration parameter and the maximum present day value the deceleration parameter will accept when $b = 0$, while the minimal value of the deceleration parameter $q$ will be observed with the interaction term given by Eq.~(\ref{eq:Q1}). If we will increase the value $b$ for the interacting models, then this increase will affect significantly on the $z_{tr}$ only for the model described by the interaction Eq.~(\ref{eq:Q1})~(bottom -- left plot of Fig.~(\ref{fig:Fig3}) ). The graphical behavior of $\Omega_{de}$ and $\Omega_{dm}$ is presented in the right column of Fig.~(\ref{fig:Fig3}) for $b=0.03$ and $b=0.05$ for the interacting models, respectively. We see that the dynamics of these two parameters carry appropriate information about the type and form of the interactions. Study of the behavior of the EoS parameter of the dark energy models showed that for appropriate values of the parameters of the models for the higher redshifts the dark energy has the phantom behavior in case of interactions given by Eq.~(\ref{eq:Q1}) and Eq.~(\ref{eq:Q4}). On the other hand, in case of non interacting dark energy model and appropriate interacting dark energy models with the interaction terms given by Eq.~(\ref{eq:Q2}) and Eq.~(\ref{eq:Q3}), the quintessence nature of the dark energy at the higher redshifts is observed. However, independent from the nature observed at the higher redshifts, during the evolution the dark energy changes its nature and at the lower redshifts we have either a quintessence universe, or a phantom universe, where the value of the EoS parameter is within the constraints coming from the new observational data. This can be seen from the left column of Fig.~(\ref{fig:Fig4}). Moreover, the top plot corresponds to the case when the interaction parameter for interacting models is $b=0.03$, while the case when $b=0.05$ corresponds to the bottom panel plot. Observed phantom nature of the EoS of the dark energy at higher redshifts has an appropriate imprint on the deceleration parameter $q$ presented in Fig.~(\ref{fig:Fig3}). The right column of Fig.~(\ref{fig:Fig4}) represents the behavior of the Hubble parameter. We can see, that for the higher redshifts consideration of the interaction, Eq.~(\ref{eq:Q1}), will provide a significant lowering of the Hubble parameter $H$, while for the lower redshifts observed difference will disappear making interacting models comparable with non interacting model. Moreover, increasing the value of $b$ will decrease the Hubble parameter at higher redshifts. The present day values of the deceleration parameter $q$, $(\omega_{de},\omega^{\prime}_{de})$ of interacting dark energy, $(r,s)$ statefinder parameters and the value of the transition redshift for some values of the interaction parameter $b$ for interactions given by Eq.~(\ref{eq:Q2}), Eq.~(\ref{eq:Q3}) and Eq.~(\ref{eq:Q4}) can be found in Table~\ref{tab:Table2}, Table~\ref{tab:Table3} and Table~\ref{tab:Table4}, respectively.       

\begin{table}[ht!]
  \centering
    \begin{tabular}{ | l | l | l | l | p{1cm} |}
    \hline
 $b$ & $q$ & $(\omega_{de},\omega^{\prime}_{de})$ & $(r,s)$ & $z_{tr}$ \\
    \hline
 $0.0 $ & $-0.522$ & $(-0.978, 0.249)$ & $(2.86, -0.61)$ & $0.682$ \\
    \hline
 $0.01$ & $-0.525$ & $(-0.981, 0.254)$ & $(2.83, -0.59)$ & $0.692$ \\
    \hline
 $0.03$ & $-0.531$ & $(-0.987, 0.264)$ & $(2.75, -0.57)$ & $0.713$ \\
    \hline
 $0.05$ & $-0.538$ & $(-0.993, 0.274)$ & $(2.69, -0.54)$ & $0.735$ \\
    \hline
 $0.07$ & $-0.544$ & $(-0.999, 0.283)$ & $(2.62, -0.52)$ & $0.757$ \\
    \hline            
    \end{tabular}
\caption{Present day values of the deceleration parameter $q$, $(\omega_{de},\omega^{\prime}_{de})$ of interacting dark energy, $(r,s)$ statefinder parameters and the value of the transition redshift $z_{tr}$ for several values of the interaction parameter $b$, when the interaction is given via Eq.~(\ref{eq:Q2}). The best fit of the theoretical results to the recent observational data has been obtained for $H_{0}=0.7$, $\alpha_{0} = 0.15$, $\alpha_{1} = 0.25$, $c = 0.75$.}
  \label{tab:Table2}
\end{table}

\begin{table}[ht!]
  \centering
    \begin{tabular}{ | l | l | l | l | p{1cm} |}
    \hline
 $b$ & $q$ & $(\omega_{de},\omega^{\prime}_{de})$ & $(r,s)$ & $z_{tr}$ \\
      \hline
 $0.0 $ & $-0.522$ & $(-0.978, 0.249)$ & $(2.86,  -0.61)$ & $0.682$ \\
          \hline
 $0.01$ & $-0.529$ & $(-0.985, 0.241)$ & $(2.863, -0.604)$ & $0.691$ \\
    \hline
 $0.03$ & $-0.544$ & $(-0.999, 0.225)$ & $(2.865, -0.596)$ & $0.708$ \\
     \hline
 $0.05$ & $-0.559$ & $(-1.02,  0.21)$ & $(2.863, -0.587)$ & $0.727$ \\
     \hline
 $0.07$ & $-0.573$ & $(-1.028, 0.196)$ & $(2.885, -0.023)$ & $0.747$ \\
     \hline            
    \end{tabular}
\caption{Present day values of the deceleration parameter $q$, $(\omega_{de},\omega^{\prime}_{de})$ of interacting dark energy, $(r,s)$ statefinder parameters and the value of the transition redshift $z_{tr}$ for several values of the interaction parameter $b$, when the interaction is given via Eq.~(\ref{eq:Q3}). The best fit of the theoretical results to the recent observational data has been obtained for $H_{0}=0.7$, $\alpha_{0} = 0.15$, $\alpha_{1} = 0.25$, $c = 0.75$.}
\label{tab:Table3}
\end{table}

\begin{table}[ht!]
  \centering
    \begin{tabular}{ | l | l | l | l | p{1cm} |}
    \hline
 $b$ & $q$ & $(\omega_{de},\omega^{\prime}_{de})$ & $(r,s)$ & $z_{tr}$ \\
      \hline
 $0.0 $ & $-0.522$ & $(-0.978, 0.249)$ & $(2.86,  -0.61)$ & $0.682$ \\
          \hline
 $0.01$ & $-0.523$ & $(-0.98,  0.255)$ & $(2.83, -0.596)$ & $0.701$ \\
    \hline
 $0.03$ & $-0.526$ & $(-0.982, 0.267)$ & $(2.77, -0.574)$ & $0.732$ \\
     \hline
 $0.05$ & $-0.528$ & $(-0.985, 0.279)$ & $(2.705, -0.553)$ & $0.771$ \\
     \hline
 $0.07$ & $-0.532$ & $(-0.987, 0.291)$ & $(2.645, -0.532)$& $0.814$ \\
     \hline            
    \end{tabular}
\caption{Present day values of the deceleration parameter $q$, $(\omega_{de},\omega^{\prime}_{de})$ of interacting dark energy, $(r,s)$ statefinder parameters and the value of the transition redshift $z_{tr}$ for several values of the interaction parameter $b$, when the interaction is given via Eq.~(\ref{eq:Q4}). The best fit of the theoretical results to the recent observational data has been obtained for $H_{0}=0.7$, $\alpha_{0} = 0.15$, $\alpha_{1} = 0.25$, $c = 0.75$.}
\label{tab:Table4}
\end{table}

\section{$Om$ and $Om3$ diagnostics with the statefinder hierarchy}\label{sec:Om}
\begin{figure}[h!]
 \begin{center}$
 \begin{array}{cccc}
\includegraphics[width=80 mm]{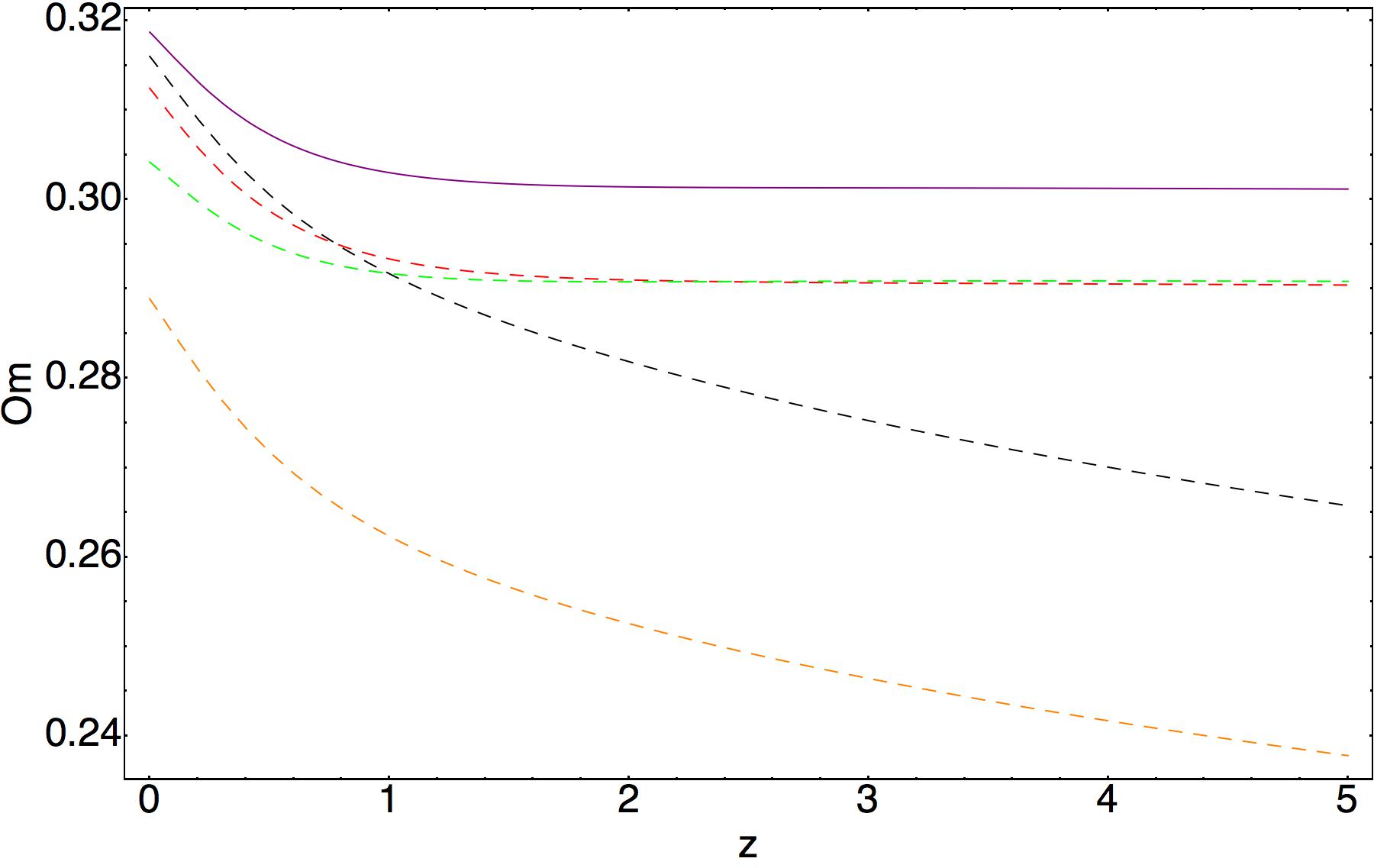} &
\includegraphics[width=80 mm]{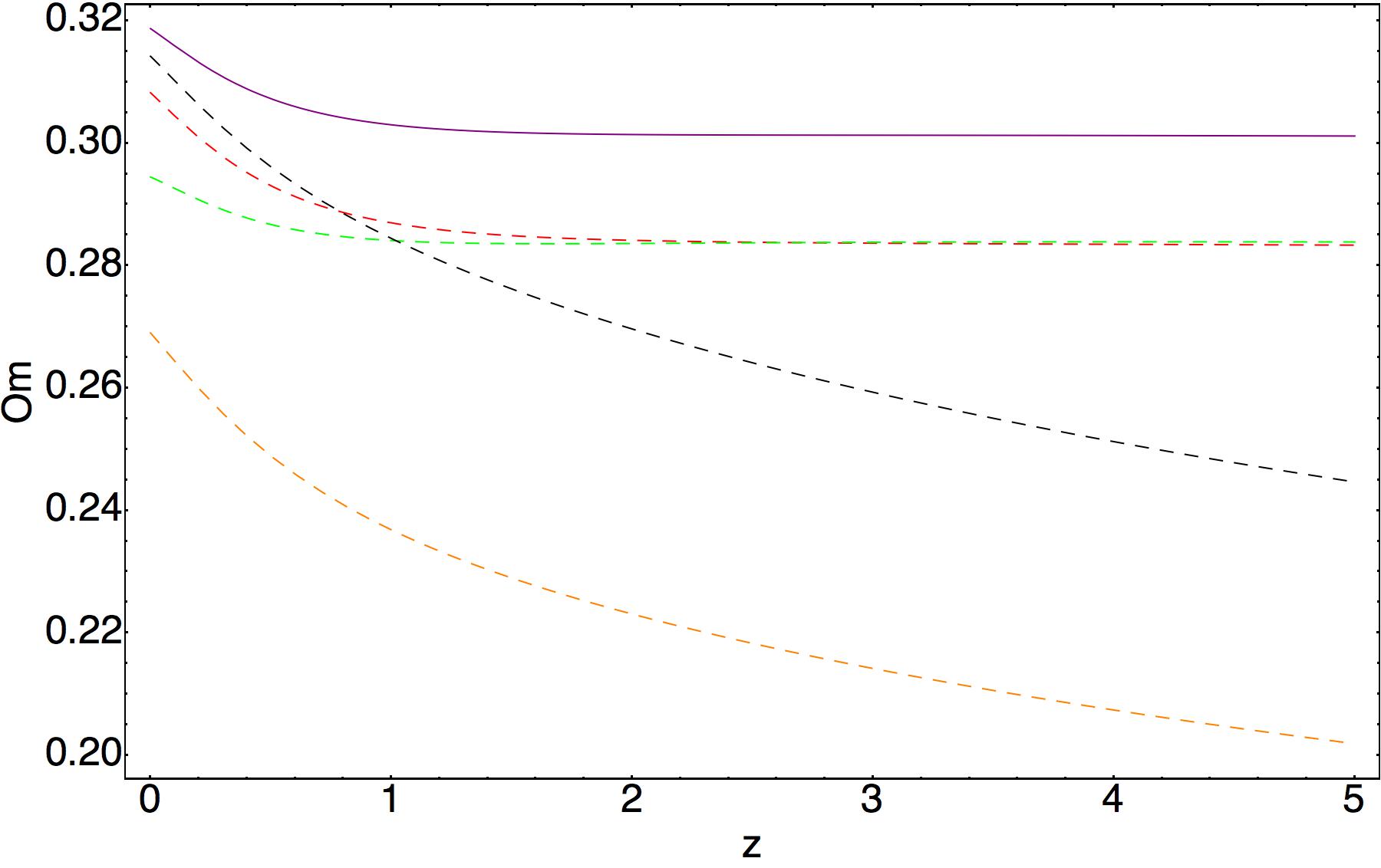} \\
 \end{array}$
 \end{center}
\caption{Graphical behavior of the $Om$ parameter against the redshift $z$. The blue curve represents non interacting model with $b = 0$, the orange curve represents the model, when the interaction is given by Eq.~(\ref{eq:Q1}), the red curve represents the model when the interaction is given by Eq.~(\ref{eq:Q2}), the green curve represents the model with the interaction given by Eq.~(\ref{eq:Q3}), while the black curve represents the model with the interaction given by Eq.~(\ref{eq:Q4}), when $H_{0}=0.7$, $\alpha_{0} = 0.15$, $\alpha_{1} = 0.25$, $c = 0.75$. The left plot corresponds to the case when $b = 0.03$. The right plot represents the case when for the interacting models $b=0.05$.}
 \label{fig:Fig5}
\end{figure}

\begin{figure}[h!]
 \begin{center}$
 \begin{array}{cccc}
\includegraphics[width=80 mm]{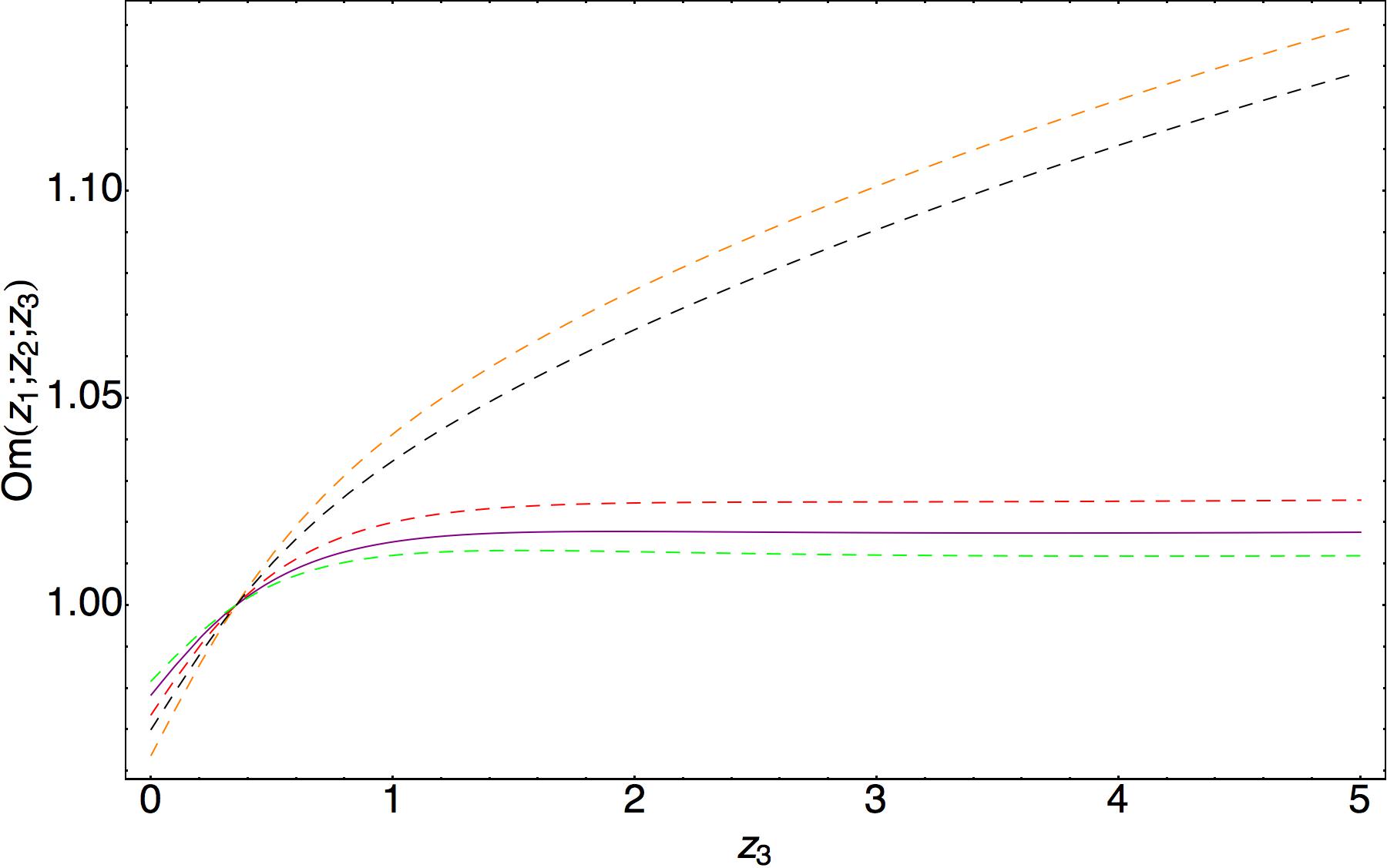} &
\includegraphics[width=80 mm]{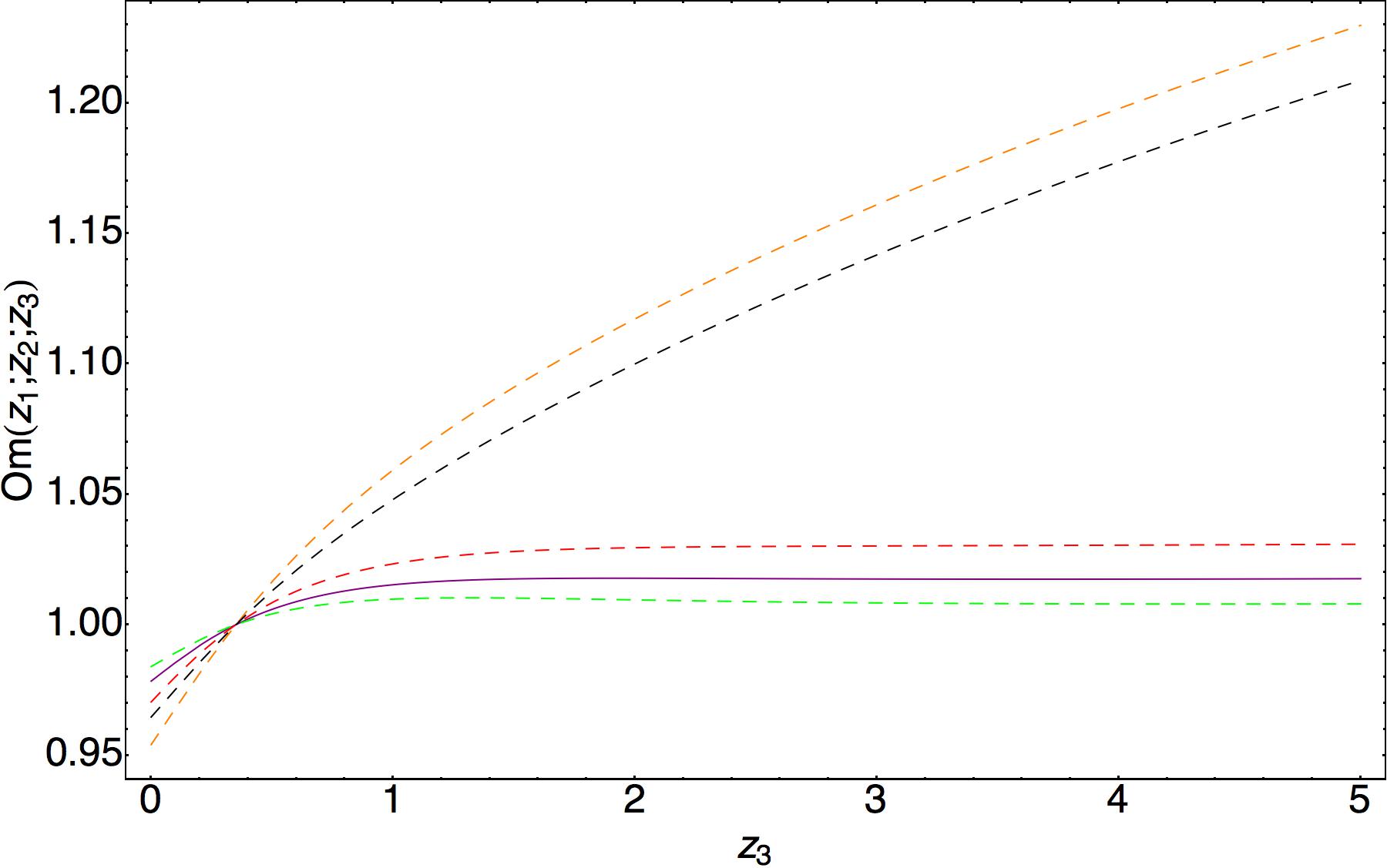} \\
 \end{array}$
 \end{center}
\caption{Graphical behavior of the $Om3$ parameters against the redshift $z$. The blue curve represents non interacting model with $b = 0$, the orange curve represents the model, when the interaction is given by Eq.~(\ref{eq:Q1}), the red curve represents the model when the interaction is given by Eq.~(\ref{eq:Q2}), the green curve represents the model with the interaction given by Eq.~(\ref{eq:Q3}), while the black curve represents the model with the interaction given by Eq.~(\ref{eq:Q4}), when $H_{0}=0.7$, $\alpha_{0} = 0.15$, $\alpha_{1} = 0.25$, $c = 0.75$. The left plot corresponds to the case when $b = 0.03$. The right plot represents the case when for the interacting models $b=0.05$. $z_{1} = 0.3$ and $z_{2} = 0.35$.}
 \label{fig:Fig6}
\end{figure}

\begin{figure}[h!]
 \begin{center}$
 \begin{array}{cccc}
\includegraphics[width=80 mm]{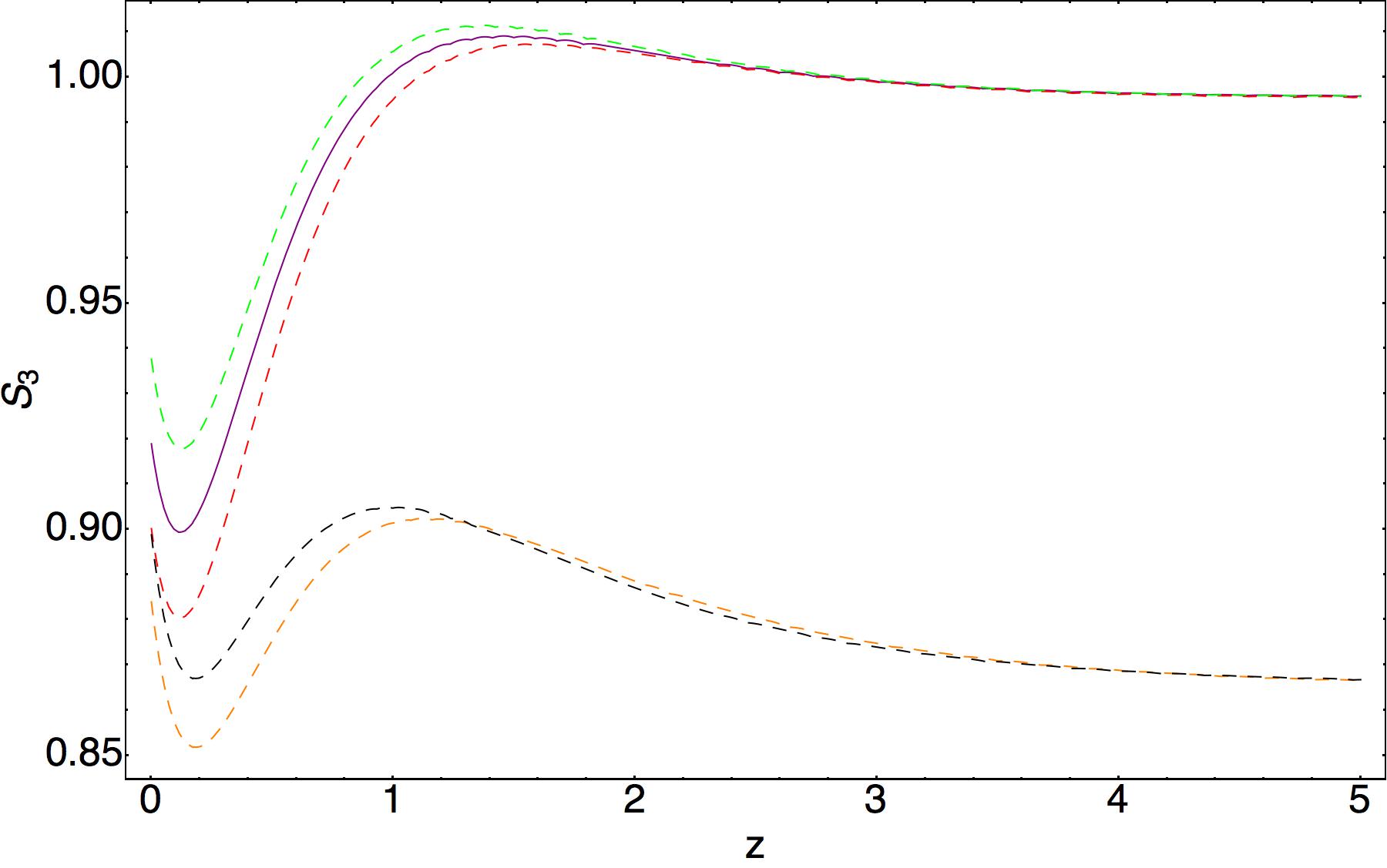} &
\includegraphics[width=80 mm]{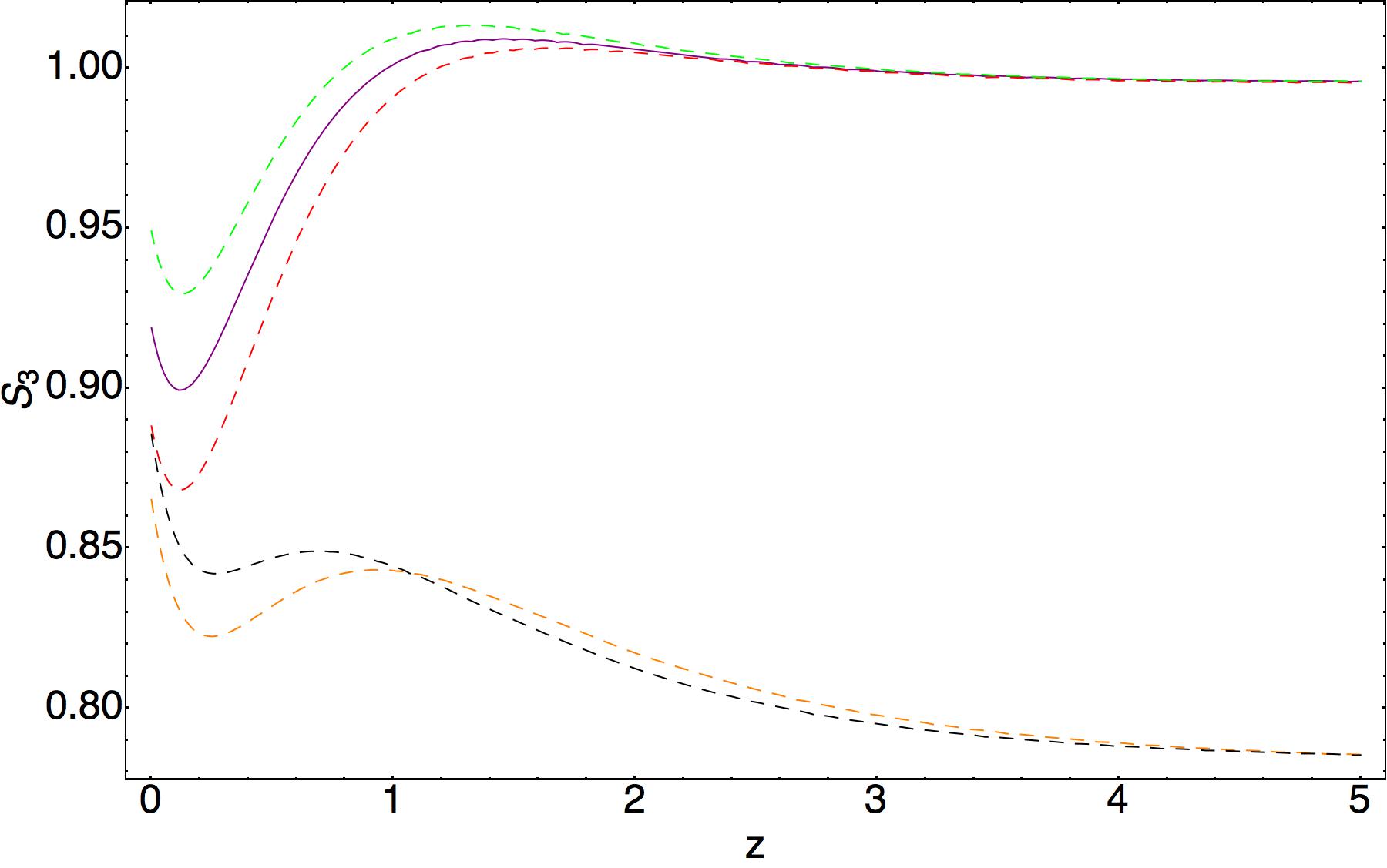} \\
 \end{array}$
 \end{center}
\caption{Graphical behavior of the $S_{3}$ parameters against the redshift $z$. The blue curve represents non interacting model with $b = 0$, the orange curve represents the model, when the interaction is given by Eq.~(\ref{eq:Q1}), the red curve represents the model when the interaction is given by Eq.~(\ref{eq:Q2}), the green curve represents the model with the interaction given by Eq.~(\ref{eq:Q3}), while the black curve represents the model with the interaction given by Eq.~(\ref{eq:Q4}), when $H_{0}=0.7$, $\alpha_{0} = 0.15$, $\alpha_{1} = 0.25$, $c = 0.75$. The left plot corresponds to the case when $b = 0.03$. The right plot represents the case when for the interacting models $b=0.05$.}
 \label{fig:Fig7}
\end{figure}

Study presented in previous section indicates possible impact of considered interactions on the behavior of the cosmological parameters. Performed study shows viability of considered models to the problems of the large scale universe, therefore additional closer look to these models is required. We already have mentioned about several analysis developed in modern cosmology and in this section we will concentrate our attention on $Om$, $Om3$ and the statefinder hierarchy analysis. From the material presented in section~\ref{s:INT} we see that statefinders use the second, third and higher order derivatives of the scale factor with respect to cosmic time whereas $Om$ relies on first order derivative alone and which combines the Hubble parameter and redshift. Consequently $Om$ is a simpler diagnostic when applied to observations. It is well known that constant behaviour of $Om$ with respect to $z$ signifies that the dark energy is a cosmological constant $\Lambda$. The positive slope of $Om$ implies that the dark energy is phantom, whereas the negative slope means that the dark energy behaves like quintessence. On the other hand, deriving $H(z)$ directly from cosmological observables in a purely model - independent and nonparameteric manner is not always an easy task. However, it is well known that $Om3$, in contrast to $Om$, is specifically tailored to be applied to baryon acoustic oscillation data and in this case, $Om3$ depends on a smaller number of cosmological observables than $Om$ making this analysis very attractive. The graphical behavior~(redshift dependent) corresponding to the $Om$ analysis can be found in Fig.~(\ref{fig:Fig5}). It is well known that $Om$ parameter for $\Lambda$CDM model is equal to $\Omega^{0}_{dm}$ and in our case $\Omega^{0}_{dm} = 0.3$, therefor this line will be taken into account to see possible departures from the standard model of cosmology. From Fig.~(\ref{fig:Fig5}) we see clear departure from $\Lambda$CDM model. Moreover, we see that for the non interacting model, and the models with interactions, Eq.~(\ref{eq:Q2}) and Eq.~(\ref{eq:Q3}), at higher redshifts the $Om$ parameter it is a constant, while at lower redshifts it is an increasing function of the redshift. On the other hand, for the models with the interactions, Eq.~(\ref{eq:Q1}) and Eq.~(\ref{eq:Q4}), the $Om$ parameter is an increasing function. Moreover, for the higher redshifts where the interacting dark energy has a phantom nature, the $Om$ parameter increases linearly. We see also, that the $Om$ analysis is a good tool to distinguish considered cosmological models from each other, too. Both plots of Fig.~(\ref{fig:Fig5}) allow to understand about differences between the models depending on the value of the interaction parameter $b$ in term of the $Om$ parameter and analysis. Three point $Om3$ analysis it is another option to have a look to suggested cosmological models. In case of this analysis for $\Lambda$CDM model the $Om3 = 1$ is taken as a reference frame. Results corresponding to $Om3$ analysis for the models are presented in Fig.~(\ref{fig:Fig6}) for two cases: $b=0.03$ represents the left plot, while the right plot represents the case where for the interacting models the interaction parameter is $b=0.05$. According to obtained results $Om3$ analysis shows that at lower redshifts there is a very tiny period in the history of the universe where suggested models marge into one and any kind of interaction will be turned off, even when at higher and lower redshifts considered models show qualitatively different behavior. $Om3$ analysis indicates clear view showing departures from $\Lambda$CDM standard model of cosmology. As we can see from Fig.~(\ref{fig:Fig6}) $Om3$ parameter is linearly decreasing $z$ dependent function for the models~(for appropriate higher redshifts) where the dark energy is a phantom dark energy. Results corresponding to the statefinder hierarchy analysis showed that $S_{3}$ in the hierarchy is well determined and provide a proper understanding of the models. Particularly, from the graphical behavior of the $S_{3}$ parameter presented in Fig.~(\ref{fig:Fig7}), we see that for a look to considered models this parameter is good enough for the lower redshifts. In this case for the higher refdshifts, as already has been demonstrated, we can use either $Om$ or $Om3$ analysis. 

\section{Thermodynamics}\label{sec:ST}
\begin{figure}[h!]
 \begin{center}$
 \begin{array}{cccc}
\includegraphics[width=80 mm]{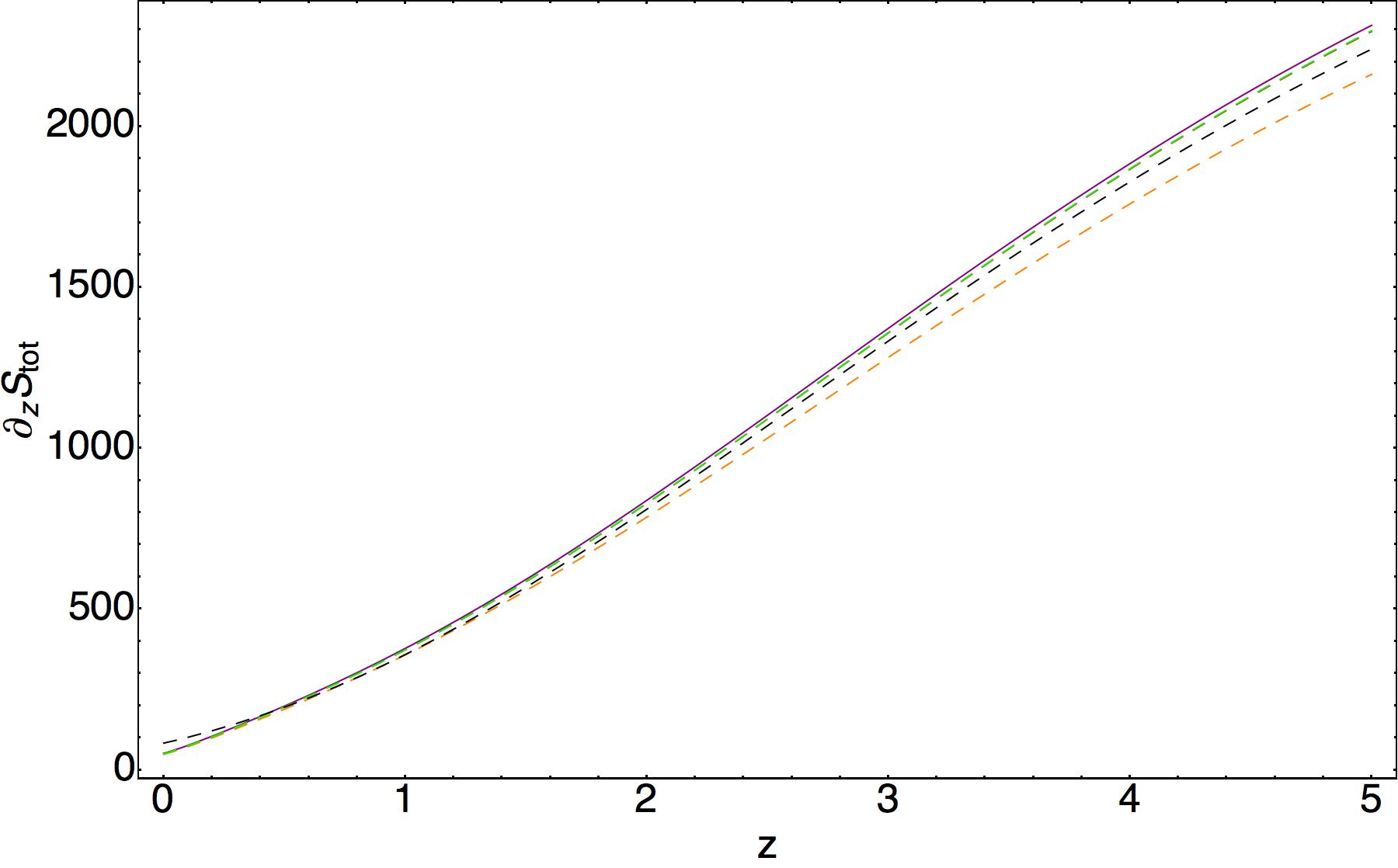} &
\includegraphics[width=80 mm]{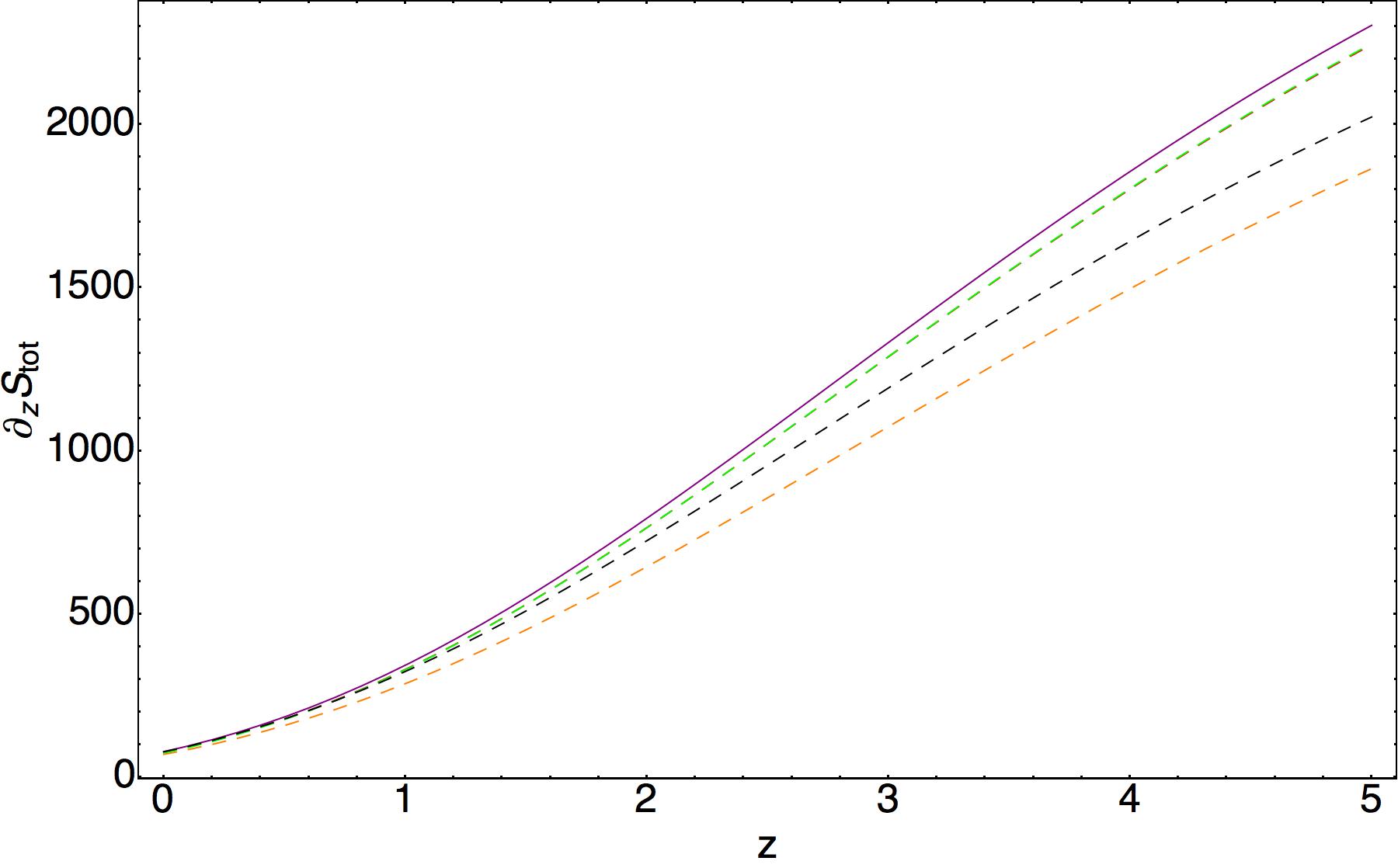} \\
 \end{array}$
 \end{center}
\caption{Graphical behavior of the dynamics of $S_{tot}$ against the redshift $z$. The blue curve represents non interacting model with $b = 0$, the orange curve represents the model, when the interaction is given by Eq.~(\ref{eq:Q1}), the red curve represents the model when the interaction is given by Eq.~(\ref{eq:Q2}), the green curve represents the model with the interaction given by Eq.~(\ref{eq:Q3}), while the black curve represents the model with the interaction given by Eq.~(\ref{eq:Q4}), when $H_{0}=0.7$, $\alpha_{0} = 0.15$, $\alpha_{1} = 0.25$, $c = 0.75$. The left plot corresponds to the case when $b = 0.01$. The right plot represents the case when for the interacting models $b=0.03$.}
 \label{fig:Fig8}
\end{figure}

Study of the dark energy and cosmological models using thermodynamics is one of the possibilities to understand the behavior of the system, which is actively discussed in recent literature. In this section we will study the question of the validity of the generalized second law of thermodynamics for considered cosmological models graphically. The generalized second law of thermodynamics in our case reads as
\begin{equation}
\dot{S}_{tot} = \dot{S}_{de} + \dot{S}_{dm} + \dot{S}_{h},  
\end{equation}
where $S_{h} = 8 \pi^{2} L^{2}$ it is the entropy associated with the horizon, while $S_{dm}$ and $S_{de}$ are the entropy associated with the dark matter and the dark energy, respectively. $\dot{}$ represents the time derivative, which in our case will be replaced by the derivative with respect to the redshift $z$. The redshift dependent dynamics of the dark energy and dark matter will be derived from the general algorithm started from the first law of thermodynamics. Particularly, we will take into account that
\begin{equation}
T dS_{i} = dE_{i} + P_{i}dV,
\end{equation} 
where $E_{i}$ it is the energy of the component and it is given as
\begin{equation}
E_{i} = \rho_{i} V,
\end{equation} 
while $V$ it is the volume of the system defined as
\begin{equation}
V = \frac{4 \pi}{3} L^{3}.
\end{equation}  
If we will take into account the model of the dark energy and the forms of the interactions considered in this paper, it is easy to obtain the form of the dynamics of the entropy of the dark energy. Particularly, if we consider the model with the interaction given by Eq.~(\ref{eq:Q1}), then for the dynamics of the entropy we will obtain the following form 

\begin{equation}
TdS_{de} = \frac{4 \pi  c^3  \dot{L}_{f}  \left(\alpha_{1} + H \left(-\sqrt{\Omega _{de}}\right)+2 \alpha_{2} L_{f}\right) \left(H \sqrt{\Omega_{de}} \left(-3 b H L_{f} \Omega_{de}+3 b H L_{f}+2\right)+2 \dot{L}_{f} (\alpha_{1}+2 \alpha_{2} L_{f})\right)}{H^4 L_{f}^2 \Omega_{de}^{3/2}},
\end{equation}
which with $b = 0$ will describe the dynamics of the entropy corresponding to the non interacting case. Following to the same reasoning it is not hard to obtain the dynamics of the entropy of the dark energy for other cosmological models considered in this paper. From the graphical behavior of the redshift dependent dynamics of the $S_{tot}$ presented in Fig.~(\ref{fig:Fig8}) we see the validity of the generalized second law of thermodynamics for considered models. Moreover, we clearly see possible imprint of considered interactions into the dynamics of the total entropy of the system. Validity of the generalized second law of thermodynamics brings additional constraints on the parameters of the models, particularly, validity of the generalized second law of thermodynamics gives a possibility to estimate an allowed upper limit on the interaction parameter as $0.01$. Combining the results obtained from the cosmographic analysis we conclude, that to have a viable cosmological model for the interaction parameter we should consider the following range $b\in[0,0.01)$, when at the higher redshifts the phantom nature of the dark energy is allowed.     
 
\section{\large{Discussion}}\label{sec:Discussion}
One of the long standing problems of modern cosmology it is the accelerated expansion of the universe. The problem can be solved either by the concept of dark energy, or by an appropriate modification of general relativity. In both directions there is an active research. However, unfortunately, existing symmetries in recent understanding do not allow us to reduce the amount of the models discussed in literature. On the other hand, the same symmetries allow to introduce very interesting ideas like interaction between dark energy and dark matter among the others. In this paper we have concentrated our attention to the phenomenological models of the large scale universe, where a specific dark energy interacts with the pressureless dark matter. Three forms of interaction considered in this work are examples of non linear interactions intensively studied in recent universe, therefore study of suggested phenomenological models will extend and complete previously considered models with the same structure of the darkness of the large scale universe. In our models we invoked a generalized holographic dark energy with a Nojiri - Odintsov cut - off and have studied the problem of the accelerated expansion of the large scale universe depends on the interaction existing inside of suggested darkness. A detailed analysis of the models and systematic comparison of the theoretical results between non interacting and interacting models have been performed. Besides the cosmographic analysis an estimation of the deceleration parameter $q$, $(\omega_{de},\omega^{\prime}_{de})$ of interacting dark energy, $(r,s)$ statefinder parameters and the value of the transition redshift $z_{tr}$ for several values of the interaction parameter $b$ for each case has been done and summarized in appropriate tables. Consideration of the generalized holographic dark energy with a Nojiri - Odintsov cut - off~(in this paper we have concentrated our attention only on a specific model) in early studies showed a possibility of the unifying of the early-time and late-time universe based on phantom cosmology. Moreover, an existing interests towards to this model of the dark energy is related to the possibility of phantom -- non-phantom transition symmetry, which appears in such a way that universe could have effectively phantom equation of state at early time as well as at late time. Generally, it is not explored yet, but we could believe that the oscillating universe may have several phantom and non-phantom phases making such models very attractive. In our study we also have seen  phantom -- non-phantom transition symmetry for early universe and for recent universe depending on the form of the interaction. We should remember, that recent constraints from the Planck 2015 satellite experiments have dramatically reduced our believe towards to the phantom large scale universe in future, however still there is such probability. Particularly, during our study we have observed, that for the values of the parameters of the model in absence of the interaction in the darkness of the large scale universe, if we have only a quintessence dark energy model, then an appropriate form of interaction can change this picture in a very interesting way. Study of the behavior of the EoS parameter of the dark energy models showed that for appropriate values of the parameters of the models for the higher redshifts the dark energy has the phantom behavior in case of interactions given by Eq.~(\ref{eq:Q1}) and Eq.~(\ref{eq:Q4}). On the other hand, in case of non interacting dark energy model and appropriate interacting dark energy models with the interaction terms given by Eq.~(\ref{eq:Q2}) and Eq.~(\ref{eq:Q3}), the quintessence nature of the dark energy at the higher redshifts is observed. However, independent from the nature observed at the higher redshifts, during the evolution the dark energy changes its nature and at the lower redshifts we have either a quintessence universe, or a phantom universe, where the value of the EoS parameter is within the constraints coming from the new observational data. The rich/different behavior of the EoS parameter of the dark energy demonstrates a possibility to determine the form of the interaction inside the darkness of the large scale universe according to recent observational data. Moreover, each form of the interaction leaving an unique imprint on the EoS parameter of the dark energy, leaves appropriate imprint on the dynamics of the other cosmological parameters, transition redshift $z_{tr}$ and on the present day values of these parameters. Taking into account a possibility to study the dark energy models via thermodynamics, we checked the validity of the generalized second law of thermodynamics for all phenomenological model considered in this paper. This allows to complete our study of the models using $Om$, $Om3$ and the stafinder hierarchy analysis, indicating that $Om$ and $Om3$ parameters are very good to see possible departures from $\Lambda$CDM standard model. Moreover, they are good tools to distinguish considered models from each other. On the other hand, from the graphical behavior of the $S_{3}$ parameter presented in Fig.~(\ref{fig:Fig7}), we see that for a look to considered models this parameter is good enough for the lower redshifts. Obtained results within considered phenomenological models motivated us to formulate possible future developments related to considered models and we hope that we can discuss obtained new results very soon elsewhere. Particularly, we have in mind to study future possible singularities which can be formed in future phantom universe and study possible impact of the interaction on this issue~\cite{Nojiri3}. Another study will be about structure formation in considered models involving spherical, ellipsoidal and triaxial collapse models, since understanding of the symmetries of the structures in our large scale universe provides additional constraints on the models/theories providing the dynamics of the background of the universe.

\newpage

\end{document}